\documentclass[pra,onecolumn,showkeys]{revtex4-2}

\usepackage[dvipsnames]{xcolor}
\usepackage{graphicx}
\usepackage{amsmath}
\usepackage{amsthm}
\usepackage{amssymb}
\usepackage{times}
\usepackage{braket}

\usepackage[T1]{fontenc}

\newtheorem{proposition}{Proposition}

\theoremstyle{definition}

\newtheorem{example}{Example}

\begin{document}

\title{Multipartite correlations in quantum collision models}

\author{Sergey N. Filippov}
\affiliation{Department of Mathematical Methods for Quantum
Technologies, Steklov Mathematical Institute of Russian Academy of
Sciences, Gubkina St. 8, Moscow 119991, Russia}

\begin{abstract}
Quantum collision models have proved to be useful for a clear and
concise description of many physical phenomena in the field of
open quantum systems: thermalization, decoherence, homogenization,
nonequilibrium steady state, entanglement generation, simulation
of many-body dynamics, quantum thermometry. A challenge in the
standard collision model, where the system and many ancillas are
all initially uncorrelated, is how to describe quantum
correlations among ancillas induced by successive system-ancilla
interactions. Another challenge is how to deal with initially
correlated ancillas. Here we develop a tensor network formalism to
address both challenges. We show that the induced correlations in
the standard collision model are well captured by a matrix product
state (a matrix product density operator) if the colliding
particles are in pure (mixed) states. In the case of the initially
correlated ancillas, we construct a general tensor diagram for the
system dynamics and derive a memory-kernel master equation.
Analyzing the perturbation series for the memory kernel, we go
beyond the recent results concerning the leading role of two-point
correlations and consider multipoint correlations (Waldenfelds
cumulants) that become relevant in the higher order stroboscopic
limits. These results open an avenue for a further analysis of
memory effects in the collisional quantum dynamics.
\end{abstract}

\keywords{collision model; repeated interactions; quantum
correlations; matrix product state; matrix product density
operator; tensor network; master equation; memory kernel}

\maketitle

\section{Introduction}

\begin{figure}
\centering
\includegraphics[width=9cm]{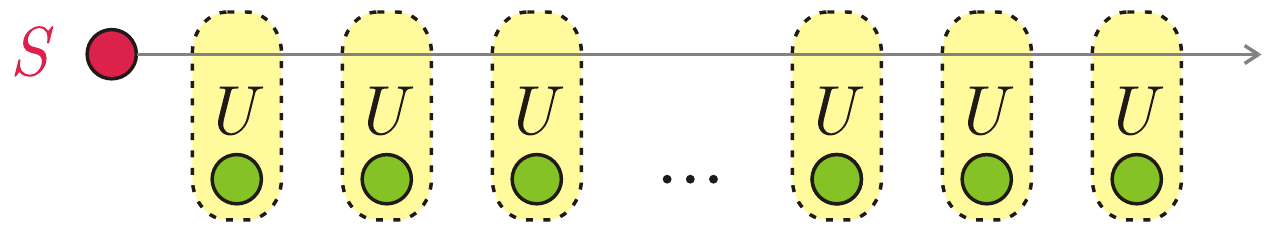}
\caption{\centering Standard collision model.} \label{figure1}
\end{figure}

The standard collision model, introduced as early as in 1963 in
Ref.~\cite{rau-1963}, considers a quantum system that sequentially
interacts with identical uncorrelated ancillary particles or
oscillator modes. Each elementary system-particle interaction
lasts for a finite period of time $\tau$ and is described by an
elementary unitary evolution operator $U$, see
Figure~\ref{figure1}. However simple this model may look like, it
(i) naturally describes the system dynamics induced by repeated
interactions, e.g., in the micromaser
theory~\cite{nachtergaele-2008}; (ii) gives an intuitively clear
picture of various phenomena such as
thermalization~\cite{scarani-2002,ziman-osid-2002},
decoherence~\cite{ziman-osid-2002,ziman-2005,grimmer-2016},
homogenization~\cite{ziman-2002,ziman-2011}, nonequilibrium steady
state~\cite{karevski-2009,filip-2021,heineken-2021}, and
entanglement
generation~\cite{heineken-2021,daryanoosh-2018,cakmak-2019}; (iii)
is amenable to analytical treatment, which makes it possible to
derive time-continuous master equations in appropriate limits on
the system-environment interaction strength and the collision
duration~\cite{attal-2006,attal-2007,vargas-2008} (in the standard
collision model, the system dynamics is Markovian and completely
positive divisible due to a past-future independence of ancillary
particles~\cite{li-2018}). Ideas of repeated interactions underlie
the discrete-time open quantum walks and their continuous-time
limit
too~\cite{attal-2012,attal-petruccione-2012,pellegrini-2014,sinayskiy-2015,liu-2017,chia-2017}.
Hence, it is no wonder that the quantum collision models are
becoming increasingly popular in quantum information, quantum
technology, and mathematical physics communities. Mysteriously,
quantum physics community and mathematical physics community know
not so much about each other and sometimes conduct a rather
isolated research on highly interrelated topics. Mathematical
physicists usually refer to the standard quantum collision model
as the repeated interaction model and treat it as a particular
model of non-equilibrium quantum statistical
mechanics~\cite{bruneau-2014}. In addition to the derivation of
master equation, the interest of mathematical physicists is also
focused on the asymptotic state in the limit of large
times~\cite{bruneau-2006,tamura-2016} and the study of random
repeated interactions~\cite{bruneau-2008,nechita-2012}. On the
other hand, quantum physicists find new applications of quantum
collision models in simulations of open quantum many-body
dynamics~\cite{purkayastha-2021,cattaneo-2021} (including
simulations on noisy intermediate-scale quantum
processors~\cite{garcia-perez-2020}), relaxation processes caused
by the dilute gas environment~\cite{filippov-2020}, quantum
thermodynamics~\cite{kosloff-2019}, and quantum
thermometry~\cite{seah-2019,strasberg-2019}. Collisional picture
of repeated interactions also takes place in quantum optics and
waveguide quantum electrodynamics, where the electromagnetic field
is represented in the form of discrete time-bin modes interacting
with a quantum
emitter~\cite{pichler-2016,guimond-2017,ciccarello-2017,gross-2018,fisher-2018,cilluffo-2020,carmele-2020,ferreira-2021,wein-2021,maffei-2022,gheri-1998,baragiola-2012,dabrowska-2020,dabrowska-2021};
however, the time-bin modes constituting the radiation field can
be correlated so that the system dynamics becomes non-Markovian
and exhibits memory effects in general. Besides the initially
correlated state of ancillary particles or
modes~\cite{gheri-1998,baragiola-2012,dabrowska-2020,dabrowska-2021,rybar-2012,filippov-2017},
memory effects in quantum collision models appear also as a result
of two-ancilla collisions in between the system-ancilla
collisions, where the latest involved ancilla interacts with the
one that would interact with the system during the next
collision~\cite{ciccarello-pra-2013,ciccarello-ps-2013,kretschmer-2016,campbell-2018}.
An alternative scenario for non-Markovian dynamics (e.g., due to
random telegraph noise) assumes that a system is composed of the
very open system under study and an auxiliary sybsystem, which
alternately interacts with a fresh reservoir ancilla and the
system under study~\cite{lorenzo-2017}. Another approach considers
repeated interactions of the system with the particles it has
already collided (including many-body
collisions)~\cite{pellegrini-2009,cilluffo-2019,taranto-2019}.
Quantum channels with memory can also be viewed in terms of
quantum collision
models~\cite{kretschmann-2005,plenio-2007,plenio-2008,rybar-2008,rybar-2009,giovannetti-2012,palma-2012,rybar-2015}.
The presented list of possible modifications for quantum collision
models is far from being complete; in this regard we refer the
interested reader to the recent review
papers~\cite{ciccarello-2022,campbell-2021}. Nonetheless, the
reader can see a great flexibility of quantum collision models to
describe a variety of physical situations in a rather simple way.

One of current challenges in the standard collision model is
related with quantum correlations among ancillas that are induced
by successive system-ancilla interactions. These correlations lead
to an advantage in the collisional quantum
thermometry~\cite{seah-2019}. However, a direct numerical
simulation of the output ancillas' state is possible for a
relatively small number $n$ of ancillas because of an
exponentially growing dimension, $d^n$, for the state of
$d$-dimensional ancillas. For instance, $d=2$ and $n \leq 12$ in
Ref.~\cite{seah-2019}. Another challenge appears if the ancillas
are initially correlated. This scenario takes place, e.g., when
the second system starts interacting with an array of ancillas
that were originally uncorrelated but previously interacted with
the first system in the standard collision
model~\cite{giovannetti-2012,palma-2012}. Alternatively, the
ancillas can represent time-bin correlated modes in the structured
electromagnetic
radiaion~\cite{pichler-2016,guimond-2017,ciccarello-2017,gross-2018,fisher-2018,cilluffo-2020,carmele-2020,ferreira-2021,wein-2021,maffei-2022,gheri-1998,baragiola-2012,dabrowska-2020,dabrowska-2021}
or particles in a correlated spin chain, e.g., spin-1 particles in
the ground state of the Affleck-Kennedy-Lieb-Tesaki (AKLT)
antiferromagnetic Hamiltonian~\cite{aklt-1987}.
Ref.~\cite{comar-2021} reports that the correlations can break
convergence of the system state to the same state of all locally
identical ancillas (such a convergence --- known as homogenization
--- would have taken place under appropriate conditions, were the
ancillas uncorrelated). Again, the exponential increase in
Hilbert-space dimension limits the numerical study in
Ref.~\cite{comar-2021} to 16 ancillas. Therefore, we face a
general problem of how to deal with correlations among ancillas
(either induced by the system or initially present).

The first goal of this paper is to represent the system-induced
correlations among ancillas (in the standard collision model) by
developing the tensor network formalism applied recently in
Ref.~\cite{filippov-2022}. The main idea behind the tensor network
representation (in the form of the matrix product
state~\cite{perez-garcia-2007,verstraete-2008,schollwock-2011,cirac-2021})
is that many $n$-partite states of $d$-dimensional ancillas
require only about $n d r^2$ complex parameters to be specified,
not $d^n$ parameters. As we show in this paper, $r$ equals the
system dimension in the standard collision model. Our second goal
is to develop the ideas of Ref.~\cite{filippov-2022} and derive a
more general master equation for the system dynamics in the
nonstandard collision model with an initially correlated
environment. The point of Ref.~\cite{filippov-2022} is that
two-point correlations among ancillas play the leading role in the
system dynamics if each elementary unitary evolution operator
slightly deviates from the identity operator. However, it may
happen that the leading contribution vanishes for a specific
interaction, and we demonstrate such an example in this paper.
Therefore, one needs to consider higher-order correlations among
ancillas and their effect on the system dynamics. We close this
gap and provide a recipe for how to derive a master equation valid
in the corresponding perturbation order for the elementary unitary
evolution operator.

\section{Tensor network notation}

Tensor network representation of quantum states is reviewed in a
number of
papers~\cite{perez-garcia-2007,verstraete-2008,schollwock-2011,cirac-2021,orus-2014,orus-2019}
and a book~\cite{montangero-2018}. Consider a pure state
$\ket{\psi}$ of $n$ particles, where each particle is associated
with a Hilbert space ${\cal H}$ of a finite dimension $d$. The
state is fully defined by $d^n$ complex numbers $C_{i_1 i_2 \ldots
i_n}$ in the decomposition
\begin{equation} \label{decomposition}
\ket{\psi} = \sum_{i_1, i_2, \ldots, i_n = 1, \ldots, d} C_{i_1
i_2 \ldots i_n} \ket{i_1} \otimes \ket{i_2} \otimes \ldots \otimes
\ket{i_n},
\end{equation}

\noindent where $\{\ket{i_k}\}_{i_k = 1, \ldots, d}$ is an
orthonormal basis in ${\cal H}$. A collection of $d^n$ complex
numbers $\{C_{i_1 i_2 \ldots i_n}\}$ can be viewed as a rank-$n$
tensor $C$ with a picture representation involving a letter
``$C$'' with $n$ legs. To distinguish the ket-vector $\ket{\psi}$
from the bra-vector $\bra{\psi}$ we add arrows to the legs,
namely, we associate outcoming arrows with ket-vectors and
incoming legs with bra-vectors.

\begin{figure}[b]
\centering
\includegraphics[width=8cm]{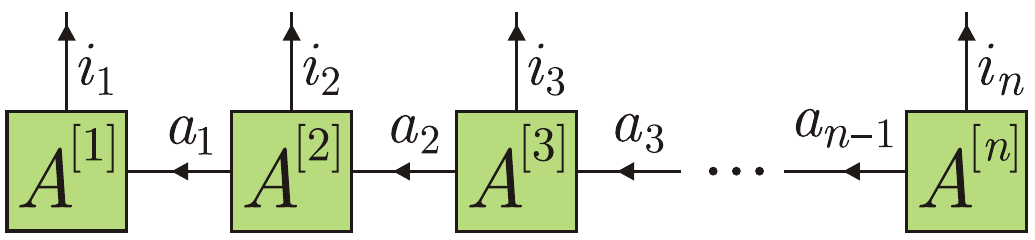}
\caption{\centering Tensor diagram for a matrix product state.}
\label{figure2}
\end{figure}

A tensor diagram concisely depicts a contraction of tensors: the
connected lines are summed over. The tensor diagram for an
$n$-partite matrix product state (MPS) with open boundary
conditions contains $n$ tensors $A^{[1]},\ldots,A^{[n]}$ connected
in a line, see Figure~\ref{figure2}. $A^{[1]}$ and $A^{[n]}$ are
rank-2 tensors with elements $A^{[1],i_1}_{a_1}$ and
$A^{[n],i_n}_{a_{n-1}}$, respectively; whereas for all
$k=2,\ldots,n-1$ the tensor $A^{[k]}$ has rank 3 and is composed
of elements $A^{[k],i_k}_{a_{k-1},a_k}$. On the other hand, if the
physical index $i_k$ is fixed, then $A^{[k],i_k}$ can be viewed as
a matrix with elements $A^{[k],i_k}_{a_{k-1},a_k}$. Similarly, if
$i_1$ and $i_n$ are fixed, then $A^{[1],i_1}$ and $A^{[n],i_n}$
can be viewed as a row-matrix and a column-matrix with matrix
elements $A^{[1],i_1}_{1,a_1}$ and $A^{[n],i_n}_{a_{n-1},1}$,
respectively. Arrows in Figure~\ref{figure2} also indicate the
order for multiplication of matrices. The contraction yields
\begin{equation} \label{eq-C}
C_{i_1 i_2 \ldots i_n} = A^{[1],i_1} A^{[2],i_2} \cdots
A^{[n],i_n},
\end{equation}

\noindent which explains the MPS name. A number $|\{a_k\}|$ of the
values that the virtual index $a_k$ can take is not related with
the physical dimension $d$ of the particles. We will refer to the
maximal number $\max_{k = 1, \ldots, n-1} |\{a_k\}|$ as the bond
dimension. Clearly, the MPS representation for a given state
$\ket{\psi}$ is not unique in general; however, the less the bond
dimension the easier the calculations and the analysis. In view of
this, the minimal bond dimension among all possible MPS
representations is called the MPS rank and denoted by $r$. The
greater $r$, the more entangled the state $\ket{\psi}$ can be with
respect to the left-right bipartitions~\cite{vidal-2003}.

\begin{figure}
\centering
\includegraphics[width=8cm]{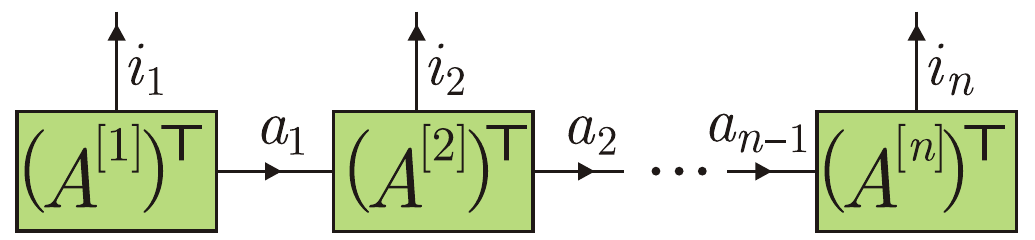}
\caption{Equivalent diagram to that in Figure~\ref{figure2}.
$\top$ denotes transposition with respect to virtual indices.}
\label{figure3}
\end{figure}

Arrows in tensor diagrams simplify their interpretation. For
instance, changing the direction of arrows from left to right in
the connecting lines in Figure~\ref{figure2}, we get the
transposed matrices $(A^{[k],i_k})^{\top}$, $k=1,\ldots,n$. The
resulting diagram is depicted in Figure~\ref{figure3}.
Nonetheless, if indices $i_1,\ldots,i_n$ are fixed, then the
c-number $C_{i_1 i_2 \ldots i_n}$ does not change because
\begin{equation} \label{eq-C-transpose}
C_{i_1 i_2 \ldots i_n} \equiv \underset{1 \times 1
\text{~matrix}}{(C_{i_1 i_2 \ldots i_n})} = (C_{i_1 i_2 \ldots
i_n})^{\top} = (A^{[n],i_n})^{\top} \cdots (A^{[2],i_2})^{\top}
(A^{[1],i_1})^{\top}.
\end{equation}

\section{Matrix product state correlations in the standard collision model}

\begin{figure}[b]
\centering
\includegraphics[width=9cm]{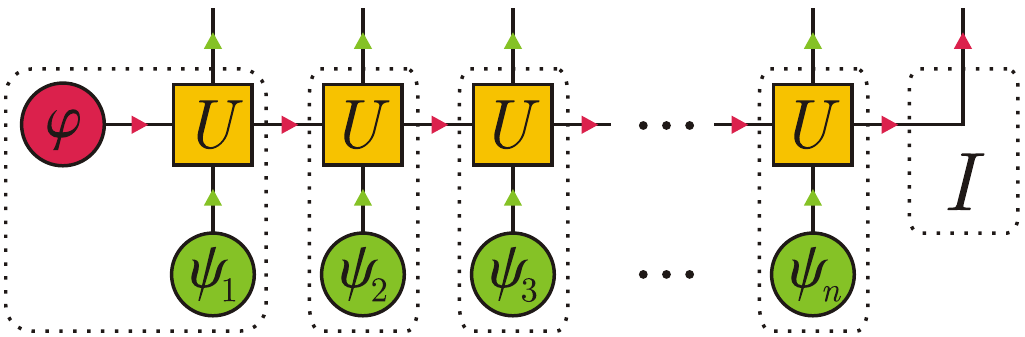}
\caption{Matrix product state of the system and ancillas induced
by collisions in the standard collision model.} \label{figure4}
\end{figure}

We begin with the simplest scenario, in which the system is
initially in a pure state $\ket{\varphi} \in {\cal H}_{S}$, ${\rm
dim}{\cal H}_S = d_S$, and the environment consists of $n$
uncorrelated ancillas in a pure state $\ket{\psi_1} \otimes
\ket{\psi_2} \otimes \ldots \otimes \ket{\psi_n} \in {\cal
H}^{\otimes n}$, ${\rm dim}{\cal H} = d$. Each elementary
collision is described by a unitary operator $U: {\cal H}_S
\otimes {\cal H} \rightarrow {\cal H}_S \otimes {\cal H}$, which
is viewed as a $4$-rank tensor. After $n$ collisions the system
and ancillas get entangled, and their composite state is given by
a tensor diagram in Figure~\ref{figure4}. As a result of $n$
collisions, we get a correlated state of $n+1$ particles: $n$
ancillas and one system particle. Dotted lines in
Figure~\ref{figure4} denote tensors that should be contracted to
get the matrix product state structure. The rightmost dotted
region depicts an identity operator $I$. Clearly, the bond
dimension equals the number of the system degrees of freedom,
$d_S$. Therefore, we can associate each virtual index $a_k$ with a
vector $\ket{a_k} \in {\cal H}_S$, so that a collection of vectors
$\{\ket{a_k}\}$ for a fixed $k$ forms an orthonormal basis in
${\cal H}_S$. The very diagram in Figure~\ref{figure4} serves as
the proof for the following result.

\begin{proposition} \label{proposition-mps}
Let the system and $n$ ancillas be initially in the pure states
$\ket{\varphi}$, $\ket{\psi_1}$, \ldots, $\ket{\psi_n}$. Then the
output state $\ket{\Psi} \in {\cal H}^{\otimes n} \otimes {\cal
H}_S$ of the system and ancillas in the standard collision model
with the elementary unitary operator $U$ adopts an MPS
representation
\begin{equation} \label{Psi}
\ket{\Psi} = \sum_{i_1, i_2, \ldots, i_n = 1}^{d} \sum_{i_{n+1} =
1}^{d_S} A^{[1],i_1} A^{[2],i_2} \cdots A^{[n],i_n}
A^{[n+1],i_{n+1}} \ket{i_1} \otimes \ket{i_2} \otimes \ldots
\otimes \ket{i_n} \otimes \ket{i_{n+1}},
\end{equation}

\noindent where $A^{[1],i_1}_{1,a_1} = \bra{a_1} \otimes \bra{i_1}
U \ket{\varphi} \otimes \ket{\psi_1}$, $A^{[k],i_k}_{a_{k-1},a_k}
= \bra{a_k} \otimes \bra{i_k} U \ket{a_{k-1}} \otimes
\ket{\psi_k}$ for all $k = 2, \ldots, n$, and
$A^{[n+1],i_{n+1}}_{a_n,1} = \delta_{a_n,i_{n+1}}$.
\end{proposition}

The result of Proposition~\ref{proposition-mps} explains the
previously known observations of Ref.~\cite{ziman-2011} that the
partial swap interactions ($U = \exp[-i g \tau \sum_{i,j}
\ket{ij}\bra{ji}]$) generate $W$-type of entanglement, whereas the
controlled unitary interactions ($U = \sum_i U_i \otimes
\ket{i}\bra{i}$) generate entanglement of the
Greenberger-Horne-Zeilinger (GHZ) type. In fact, both $W$ and GHZ
states of many qubits are particular forms of the matrix product
states with the bond dimension
2~\cite{perez-garcia-2007,verstraete-2008,schollwock-2011,cirac-2021,orus-2014,orus-2019}.

\begin{example}
Let the system and ancillas be qubits. The system is initially in
the excited state $\ket{\varphi} = \ket{\uparrow}$. Each ancilla
is initially in the ground state, i.e., $\ket{\psi_k} =
\ket{\downarrow}$ for all $k$. Consider the energy exchange
unitary $U = \exp[g \tau
(\ket{\downarrow\uparrow}\bra{\uparrow\downarrow} -
\ket{\uparrow\downarrow}\bra{\downarrow\uparrow})]$. Then
Proposition~\ref{proposition-mps} yields
\begin{eqnarray*}
A^{[1],\downarrow} = \left(%
\begin{array}{cc}
  0 & \cos g\tau \\
\end{array}%
\right), \quad A^{[k],\downarrow} = \left(%
\begin{array}{cc}
  1 & 0 \\
  0 & \cos g\tau \\
\end{array}%
\right) \text{~for~} k
= 2, \ldots, n, \quad A^{[n+1],\downarrow} = \left(%
\begin{array}{c}
  1 \\
  0 \\
\end{array}%
\right), && \\
A^{[1],\uparrow} = \left(%
\begin{array}{cc}
  \sin g\tau & 0 \\
\end{array}%
\right), \quad A^{[k],\uparrow} = \left(%
\begin{array}{cc}
  0 & 0 \\
  \sin g\tau & 0 \\
\end{array}%
\right) \text{~for~} k
= 2, \ldots, n, \quad A^{[n+1],\uparrow} = \left(%
\begin{array}{c}
  0 \\
  1 \\
\end{array}%
\right). &&
\end{eqnarray*}

\noindent Note that the matrix $A^{[k],\uparrow}$ is nilpotent,
i.e., the product of the matrix with itself is equal to a null
matrix. For this reason $A^{[k],\uparrow} A^{[k+1],\uparrow} = 0$,
which means that after interactions the adjacent ancillas cannot
be in the state $\ket{\uparrow}$. Similarly, $A^{[k],\uparrow}
(\prod_{l=1}^{m-1} A^{[k+l],\downarrow}) A^{[k+m],\uparrow} = 0$,
which shows that any two ancillas cannot simultaneously occupy the
state $\ket{\uparrow}$. The system and ancillas are finally in the
$W$-like state
\begin{equation*}
\ket{\Psi} = \sum_{k=0}^{n-1} \cos^k g\tau \sin g\tau \ket{
\underbrace{\downarrow \ldots \downarrow}_{k} \uparrow
\underbrace{\downarrow \ldots \downarrow}_{n-k} } + \cos^n g\tau
\ket{ \underbrace{\downarrow \downarrow \ldots \downarrow
\downarrow}_{n} \uparrow }.
\end{equation*}
\hfill$\triangle$\end{example}

The explicit relation between the unitary operator $U$ and tensors
$A^{[k]}$, which we establish in
Proposition~\ref{proposition-mps}, enables one to approach the
quantum engineering problem too. Suppose one wants to create an
entangled state $\ket{\Psi'}$ of $n$ particles that adopts a
matrix product representation with the bond dimension $r$. Then
one needs to take an $r$-dimensional quantum system and let it
sequentially interact with the initially uncorrelated particles.
Finally, one performs a projective measurement on the system in
the basis $\{\ket{i_{n+1}}\}$ to get rid of its degrees of
freedom. The resulting state of $n$ particles is $\braket{i'_{n+1}
| \Psi}$, where $i'_{n+1}$ is the measurement outcome and
$\ket{\Psi}$ is given by Eq.~\eqref{Psi}. The unitary operator $U$
should be optimized in such a way as to maximize the overlap
$|(\bra{\Psi'} \otimes \bra{i'_{n+1}}) \ket{\Psi} |^2$. Clearly,
each collision could be described by its own unitary operator,
then one should replace $U \rightarrow U_k$ in the formula for
$A^{[k]}$ in Proposition~\ref{proposition-mps}. Numerical tools
for optimization over many unitary operators $\{U_k\}_{k=1}^n$ are
presented, e.g., in
Refs.~\cite{luchnikov-qgopt-2021,luchnikov-2021}.

\begin{figure}[b]
\centering
\includegraphics[width=14cm]{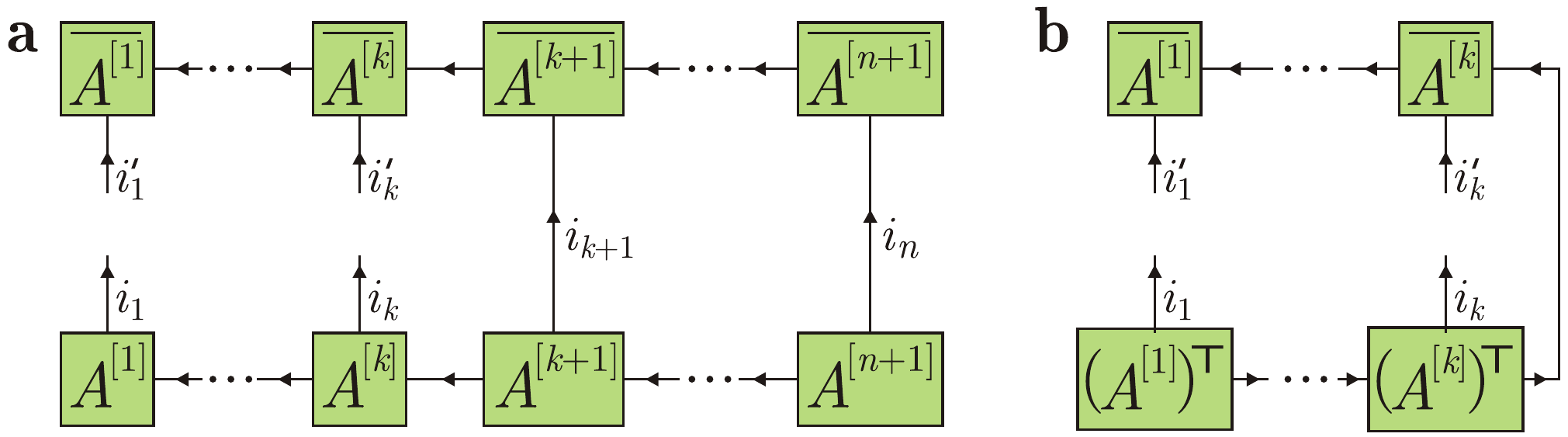}
\caption{(\textbf{a}) Tensor diagram for the reduced density
operator $\varrho_{1 \ldots k}$. Overline denotes complex
conjugation. (\textbf{b}) Simplified tensor diagram for
$\varrho_{1 \ldots k}$ due to the right normalization condition.}
\label{figure5}
\end{figure}

Let us consider entanglement of the state $\ket{\Psi}$ with
respect to a bipartition into ancillas $1,\ldots,k$ on one side
and ancillas $k+1,\ldots,n$ and the system on the other side,
i.e., the left-right bipartition with the boundary in between the
ancillas $k$ and $k+1$. Entanglement of a pure state with respect
to a bipartition is quantified by the entanglement entropy that
equals the von Neumann entropy of either reduced density operator,
$S(\varrho_{1 \ldots k}) = S (\varrho_{k+1 \ldots n S})$, where
$S(\varrho) = - {\rm tr}[\varrho \log_2 \varrho]$. The reduced
density operator $\varrho_{1 \ldots k} = {\rm tr}_{k+1,\ldots,n+1}
\ket{\Psi}\bra{\Psi}$ for $k$ ancillas is presented in the form of
a tensor diagram in Figure~\ref{figure5}a.

The tensor diagram in Figure~\ref{figure5}a gets simpler if we
take into account the following important property (referred to as
the right-normalization condition~\cite{schollwock-2011}):
\begin{equation} \label{right-normalization}
\sum_{i_m,a_m} A^{[m],i_m}_{a_{m-1},a_m}
\overline{A^{[m],i_m}_{a_{m},a'_{m-1}}} =
\delta_{a_{m-1},a'_{m-1}} \Leftrightarrow \sum_{i_m} A^{[m],i_m}
(A^{[m],i_m})^{\dag} = I \text{~for~all~} m = 1, \ldots, n+1.
\end{equation}

\noindent Here, the overline denotes the complex conjugation and
$\dag$ denotes the Hermitian conjugation. In fact, if $m=n+1$,
then $\sum_{i_{n+1}} A^{[n+1],i_{n+1}}_{a_n,1}
\overline{A^{[n+1],i_{n+1}}_{1,a'_n}} = \delta_{a_n a'_n}$ because
$A^{[n+1],i_{n+1}}_{a_n,1} = \delta_{a_n,i_{n+1}}$ by
Proposition~\ref{proposition-mps}. If $m=2,\ldots,n$, then
Proposition~\ref{proposition-mps} implies

\begin{eqnarray}
\sum_{i_m,a_m} A^{[m],i_m}_{a_{m-1},a_m}
\overline{A^{[m],i_m}_{a_{m},a'_{m-1}}} &=& \sum_{i_m,a_m}
\overline{A^{[m],i_m}_{a_{m},a'_{m-1}}} A^{[m],i_m}_{a_{m-1},a_m}
\nonumber\\
&=& \sum_{i_m,a_m} \bra{a'_{m-1}} \otimes \bra{\psi_m} U^{\dag}
(\ket{a_m} \otimes \ket{i_m}) (\bra{a_m} \otimes \bra{i_m}) U
\ket{a_{m-1}} \otimes \ket{\psi_m} \nonumber\\
&=& \bra{a'_{m-1}} \otimes \bra{\psi_m} \underbrace{U^{\dag}
U}_{I} \ket{a_{m-1}}
\otimes \ket{\psi_m} \nonumber\\
&=& \braket{a'_{m-1} | a_{m-1}} \braket{\psi_m | \psi_m} \nonumber\\
&=& \delta_{a_{m-1},a'_{m-1}}.
\end{eqnarray}

\noindent If $m=1$, then we deal with dummy indices $a_0 = a'_0 =
1$ and $\sum_{i_1,a_1} A^{[1],i_1}_{1,a_1}
\overline{A^{[1],i_1}_{a_1,1}} = \braket{\varphi|\varphi}
\braket{\psi_1 | \psi_1} = 1$. Hence, we have proved the following
result.

\begin{proposition} \label{proposition-right-normalization}
MPS $\ket{\Psi}$ in Eq.~\eqref{Psi} satisfies the
right-normalization condition~\eqref{right-normalization}.
\end{proposition}

An MPS satisfying the right normalization condition is also called
right-canonical~\cite{schollwock-2011}. The advantage of the
right-canonical form is that the partial trace over rightmost
particles corresponds to a single connecting line in the tensor
diagram, see Figure~\ref{figure5}b. Indeed,
Eq.~\eqref{right-normalization} is equivalent to $\sum_{i_m}
(A^{[m],i_m})^{\top} \overline{A^{[m],i_m}} = I$, which is exactly
the vertical connecting line in Figure~\ref{figure5}b. Physically,
the reduced density operator for $k$ ancillas does not depend on
future system collisions with other ancillas that happen after
time $k\tau$.

Entanglement entropy $E(\Psi)$ of the state $\ket{\Psi}$ with
respect to the cut in between the ancillas $k$ and $k+1$ reads
\begin{eqnarray}
E(\Psi) &=& S(\varrho_{1 \ldots k}) \nonumber\\
&=& S\left( \sum_{i_1,\ldots,i_k,i'_1,\ldots,i'_k}
\overline{A^{[1],i_1}} \cdots \overline{A^{[k],i_k}}
(A^{[k],i_k})^{\top} \cdots (A^{[1],i_1})^{\top}
\ket{i_1}\bra{i'_1} \otimes \cdots \ket{i_k}\bra{i'_k}\right)
\nonumber\\
&=& S\left( \sum_{i_1,\ldots,i_k,i'_1,\ldots,i'_k} A^{[1],i_1}
\cdots A^{[k],i_k} (A^{[k],i_k})^{\dag} \cdots
(A^{[1],i_1})^{\dag} \ket{i_1}\bra{i'_1} \otimes \cdots
\ket{i_k}\bra{i'_k}\right).
\end{eqnarray}

\noindent Note that $E(\Psi) \leq \log d_S$ because $d_S$ is an
upper bound for the Schmidt rank of $\ket{\Psi}$.

\section{Generalization to mixed states of the system and ancillas}

\begin{figure}

\centering
\includegraphics[width=18cm]{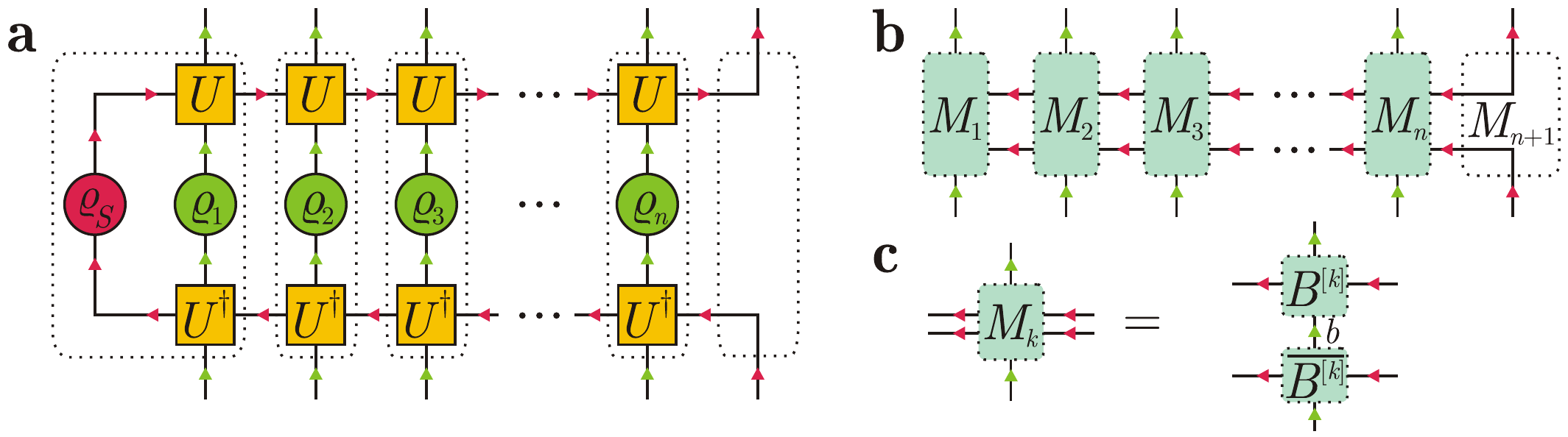}

\caption{(\textbf{a}) Tensor diagram for
Eq.~\eqref{output-state-scm}. (\textbf{b}) Matrix product density
operator. (\textbf{c}) Tensor decomposition guaranteeing positive
semidefiniteness of the matrix product density operator.}
\label{figure6}
\end{figure}

Let us consider the standard collision model, where the system and
ancillas are generally mixed. This scenario is especially relevant
to the task of quantum
thermometry~\cite{seah-2019,strasberg-2019}. The initial state of
the system is given by the density operator $\varrho_S$. The
initial state of $n$ ancillas is given by a factorized density
operator $\bigotimes_{k=1}^n \varrho_k$. Collisional dynamics with
the elementary unitary operator $U$ drives the system and ancillas
to the state
\begin{equation} \label{output-state-scm}
 U_{Sn} \cdots U_{S1}(\varrho_S \otimes \varrho_1 \otimes \ldots \otimes
\varrho_n) U_{S1}^{\dag} \cdots U_{Sn}^{\dag},
\end{equation}

\noindent where the subscript $Sk$ in the notation $U_{Sk}$ means
that $U$ nontrivially acts on the system and the $k$-th ancilla. A
tensor diagram for Eq.~\eqref{output-state-scm} is presented in
Figure~\ref{figure6}a.

\noindent Dotted regions in Figure~\ref{figure6}a show tensor
contractions or tensor combinations that effectively lead to the
equivalent tensor diagram depicted in Figure~\ref{figure6}b. Note,
however, that the arrows in the upper horizontal line in
Figures~\ref{figure6}a and \ref{figure6}b are different. The
operator in Figure~\ref{figure6}b reads
\begin{equation} \label{mpdo}
\varrho_{1 \ldots n S}(n\tau) =
\sum_{i_1,\ldots,i_n,i_{n+1},i'_1,\ldots,i'_n,i'_{n+1}} M_1^{i_1
i'_1} \cdots M_n^{i_n i'_n} M_{n+1}^{i_{n+1} i'_{n+1}} \ket{i_1
\ldots i_n i_{n+1}} \bra{i'_1 \ldots i'_n i'_{n+1}}.
\end{equation}

\noindent Here, $M_1$ and $M_{n+1}$ are rank-4 tensors, whereas
$M_k$ is a rank-6 tensor for all $k=2,\ldots,n$. If indices $i_1$
and $i'_1$ are fixed, then we treat $M_1^{i_1 i'_1}$ as a
row-matrix with elements $(M_1^{i_1 i'_1})_{11,a_1 a'_1}$.
Similarly, if indices $i_{n+1}$ and $i'_{n+1}$ are fixed, then we
treat $M_{n+1}^{i_{n+1} i'_{n+1}}$ as a column-matrix with
elements $(M_{n+1}^{i_{n+1} i'_{n+1}})_{a_{n+1} a'_{n+1},11}$. If
$k \in (2,\ldots,n)$ and indices $i_k$ and $i'_k$ are fixed, then
we treat $M_k^{i_k i'_k}$ as a matrix with elements $(M_k^{i_k
i'_k})_{a_{k-1} a'_{k-1},a_k a'_k}$, i.e., $a_{k-1} a'_{k-1}$ is a
row multiindex and $a_k a'_k$ is a column multiindex. The explicit
expressions for $M$-tensors follow from Figures~\ref{figure6}a and
\ref{figure6}b and read

\begin{eqnarray}
&& (M_1^{i_1 i'_1})_{11,a_1 a'_1} = \bra{a_1} \otimes \bra{i_1} U
(\varrho_S \otimes \varrho_1) U^{\dag} \ket{a'_1} \otimes
\ket{i'_1}, \label{M-through-rho-1} \\
&& (M_k^{i_k i'_k})_{a_{k-1} a'_{k-1},a_k a'_k} = \bra{a_k}
\otimes \bra{i_k} U (\ket{a_{k-1}}\bra{a'_{k-1}} \otimes
\varrho_k) U^{\dag} \ket{a'_k} \otimes
\ket{i'_k}, \qquad k=2,\ldots,n, \\
&& (M_{n+1}^{i_{n+1} i'_{n+1}})_{a_{n+1} a'_{n+1},11} =
\delta_{i_{n+1},a_{n+1}} \delta_{i'_{n+1},a'_{n+1}}.
\label{M-through-rho-3}
\end{eqnarray}

Let $\varrho_S = \sum_l \lambda_S^l
\ket{\varphi_S^l}\bra{\varphi_S^l}$ be the spectral decomposition
for the system initial state. Let $\varrho_k = \sum_m \lambda_k^m
\ket{\psi_k^m}\bra{\psi_k^m}$ be the spectral decomposition for
the initial state of the $k$-th ancilla. Then one readily obtains
the representation

\begin{eqnarray}
&& M_1^{i_1 i'_1} = \sum_{lm} B^{[1],i_1}_{lm} \otimes
\overline{B^{[1],i'_1}_{lm}}, \qquad (B^{[1],i_1}_{lm})_{1,a_1} =
\sqrt{\lambda_S^l \lambda_k^m} \bra{a_1} \otimes \bra{i_1} U
\ket{\varphi_S^l}
\otimes \ket{\psi_k^m}, \label{M-through-BB-1} \\
&& M_k^{i_k i'_k} = \sum_m B^{[k],i_k}_m \otimes
\overline{B^{[k],i'_k}_m}, \qquad (B^{[k],i_k}_m)_{a_{k-1},a_k} =
\sqrt{\lambda_k^m} \bra{a_k} \otimes \bra{i_k} U \ket{a_{k-1}}
\otimes \ket{\psi_k^m}, \qquad k=2,\ldots,n, \qquad \label{M-through-BB} \\
&& M_{n+1}^{i_{n+1} i'_{n+1}} =  B^{[n+1],i_{n+1}} \otimes
\overline{B^{[n+1],i_{n+1}}}, \qquad (B^{[n+1],i_{n+1}})_{a_n,1} =
\delta_{a_n,i_{n+1}}. \label{M-through-BB-3}
\end{eqnarray}

\noindent Tensor diagram for Eq.~\eqref{M-through-BB} is depicted
in Figure~\ref{figure6}c. We see that for any $k \in
(1,2,\ldots,n,n+1)$ the decomposition $M_k^{i_k i'_k} =
\sum_{b=1}^D B^{[k],i_k}_b \otimes \overline{B^{[k],i'_k}_b}$
takes place, with $b = (lm)$ and $D \leq d_S d$ if $k=1$, $b = m$
and $D \leq d$ if $k \in (2,\ldots,n)$, and $b = D = 1$ if
$k=n+1$. Tensor diagrams in Figures~\ref{figure6}b and
\ref{figure6}c define the so-called matrix product density
operator (MPDO)~\cite{verstraete-2004,zwolak-2004}, which is
automatically Hermitian and positive semidefinite. MPDOs are
successfully used to study the dissipative dynamics and the Gibbs
states of one-dimensional quantum
chains~\cite{verstraete-2004,zwolak-2004,chen-2020,bondarenko-2021}.
Among other questions, Ref.~\cite{bondarenko-2021} addresses an
important question how to prepare MPDO states experimentally. Our
results show one more method to prepare an MPDO state via the
standard collision model. In our construction, the MPDO is right
canonical, i.e., it additionally satisfies the right-normalization
condition
\begin{equation} \label{right-normalization-mpdo}
\sum_{i_k,a_k} (M_k^{i_k i_k})_{a_{k-1} a'_{k-1},a_k a_k} =
\delta_{a_{k-1} a'_{k-1}} \quad \Leftrightarrow \quad \sum_{i_k,b}
B^{[k],i_k}_b (B^{[k],i_k}_b)^{\dag} = I.
\end{equation}

\noindent Eq.~\eqref{right-normalization-mpdo} mathematically
shows independence of the reduced density operator $\varrho_{1
\ldots k}(k\tau)$ for $k$ ancillas from future collisions at times
$t > k\tau$. The results of this section are summarized as
follows.

\begin{proposition} \label{proposition-mpdo}
The standard collision model with initially mixed states of the
system ($\varrho_S$) and $n$ ancillas
($\varrho_1,\ldots,\varrho_n$) produces a right-canonical MPDO
\eqref{mpdo} with elementary tensors given by Eqs.
\eqref{M-through-rho-1}--\eqref{M-through-rho-3} and
\eqref{M-through-BB-1}--\eqref{M-through-BB-3}.
\end{proposition}

The main benefit of the constructed MPDO representation is that it
exploits only $d_S^2 d^2 + (n-1) d_S^4 d^2 \leq n d_S^4 d^2$
parameters instead of $d_S^2 d^{2n}$ parameters needed for a
description of a general state of the system and $n$ ancillas. In
other words, computational resources scale linearly (not
exponentially) with the number of ancillas if one uses the MPDO
representation. This fact opens an avenue for further numerical
studies in the collisional quantum
thermometry~\cite{seah-2019,strasberg-2019}. If the system
interacts with a thermal reservoir in between the collisions with
ancillas, one can readily include such a system-reservoir
interaction in the tensor network representation in the form of a
quantum channel~\cite{wood-2015}.

\begin{example}
Let the system and ancillas be qubits. The system is initially in
the excited state $\ket{\varphi} = \ket{\uparrow}$. Each ancilla
is initially in the Gibbs state
\begin{equation*}
\varrho_k = \frac{1}{1+\exp[(E_{\downarrow} - E_{\uparrow})/k_{\rm
B} T]} \ket{\downarrow}\bra{\downarrow} +
\frac{1}{1+\exp[(E_{\uparrow} - E_{\downarrow})/k_{\rm B} T]}
\ket{\uparrow}\bra{\uparrow},
\end{equation*}

\noindent where $k_{\rm B}$ is the Boltzmann constant, $T$ is the
temperature, $E_{\uparrow}$ and $E_{\downarrow}$ are the energy
levels for the ancilla states $\ket{\uparrow}$ and
$\ket{\downarrow}$, respectively. Consider the energy exchange
unitary $U = \exp[g \tau
(\ket{\downarrow\uparrow}\bra{\uparrow\downarrow} -
\ket{\uparrow\downarrow}\bra{\downarrow\uparrow})]$. After $n$
collisions, the mixed state of the system and ancillas is fully
described by a right-canonical MPDO with $D=2$. The explicit form
for this MPDO is given by Proposition~\ref{proposition-mpdo} and
reads

\begin{eqnarray*}
&& B_1^{[1],\downarrow} = \frac{1}{\sqrt{1+\exp[(E_{\downarrow} - E_{\uparrow})/k_{\rm B} T]}} \left(%
\begin{array}{cc}
  0 & \cos g\tau \\
\end{array}%
\right), \quad B_2^{[1],\downarrow} = \left(%
\begin{array}{cc}
  0 & 0 \\
\end{array}%
\right), \\
&& B_1^{[1],\uparrow} = \frac{1}{\sqrt{1+\exp[(E_{\downarrow} - E_{\uparrow})/k_{\rm B} T]}} \left(%
\begin{array}{cc}
  \sin g\tau & 0 \\
\end{array}%
\right), \quad B_2^{[1],\uparrow} =
\frac{1}{\sqrt{1+\exp[(E_{\uparrow} -
E_{\downarrow})/k_{\rm B} T]}} \left(%
\begin{array}{cc}
  0 & 1 \\
\end{array}%
\right), \\
&& B_1^{[k],\downarrow} = \frac{1}{\sqrt{1+\exp[(E_{\downarrow} - E_{\uparrow})/k_{\rm B} T]}} \left(%
\begin{array}{cc}
  1 & 0 \\
  0 & \cos g\tau \\
\end{array}%
\right), \quad B_2^{[k],\downarrow} =
\frac{1}{\sqrt{1+\exp[(E_{\uparrow} -
E_{\downarrow})/k_{\rm B} T]}} \left(%
\begin{array}{cc}
  0 & -\sin g\tau \\
  0 & 0 \\
\end{array}%
\right), \\
&& B_1^{[k],\uparrow} = \frac{1}{\sqrt{1+\exp[(E_{\downarrow} - E_{\uparrow})/k_{\rm B} T]}} \left(%
\begin{array}{cc}
  0 & 0 \\
  \sin g\tau & 0 \\
\end{array}%
\right), \quad B_2^{[k],\uparrow} =
\frac{1}{\sqrt{1+\exp[(E_{\uparrow} -
E_{\downarrow})/k_{\rm B} T]}} \left(%
\begin{array}{cc}
  \cos g\tau & 0 \\
  0 & 1 \\
\end{array}%
\right), \\
&& B^{[n+1],\downarrow} = \left(%
\begin{array}{c}
  1 \\
  0 \\
\end{array}%
\right) \quad B^{[n+1],\uparrow} = \left(%
\begin{array}{c}
  0 \\
  1 \\
\end{array}%
\right),
\end{eqnarray*}

\noindent where $k = 2, \ldots, n$. \hfill$\triangle$\end{example}

\section{Collision model with a generally correlated state of ancillas}

Let us consider a more complicated collision model, in which
ancillas are initially correlated. Surprisingly enough, any pure
state of $n$ ancillas adopts an MPS
representation~\cite{perez-garcia-2007,verstraete-2008,schollwock-2011,cirac-2021,orus-2014,orus-2019}.
However, the MPS rank for a generally correlated state grows
exponentially with $n$. On the other hand, many important states
of correlated ancillas such as few-photon
wavepackets~\cite{gheri-1998,baragiola-2012,dabrowska-2020,dabrowska-2021},
artificial photonic tensor network
states~\cite{guimond-2017,dhand-2018,lubash-2018,istrati-2020,besse-2020,tiurev-2020,wei-2021},
and ground states of gapped one-dimensional local Hamiltonians for
the spin chains~\cite{dalzell-2019} are described by MPSs with a
low MPS rank. As the state of ancillas is mixed in general, we
exploit the MPDO formalism. We pay little attention to the rank of
decomposition as our further goal is to reveal the effect of
ancillas' correlations on the system dynamics. Note that the
correlations can be either quantum (genuinely entangled ancillas)
or classical (fully separable state of ancillas); however, both
types strongly affect the system dynamics (see an example in
Ref.~\cite{rybar-2012}).

\begin{figure}
\centering
\includegraphics[width=14cm]{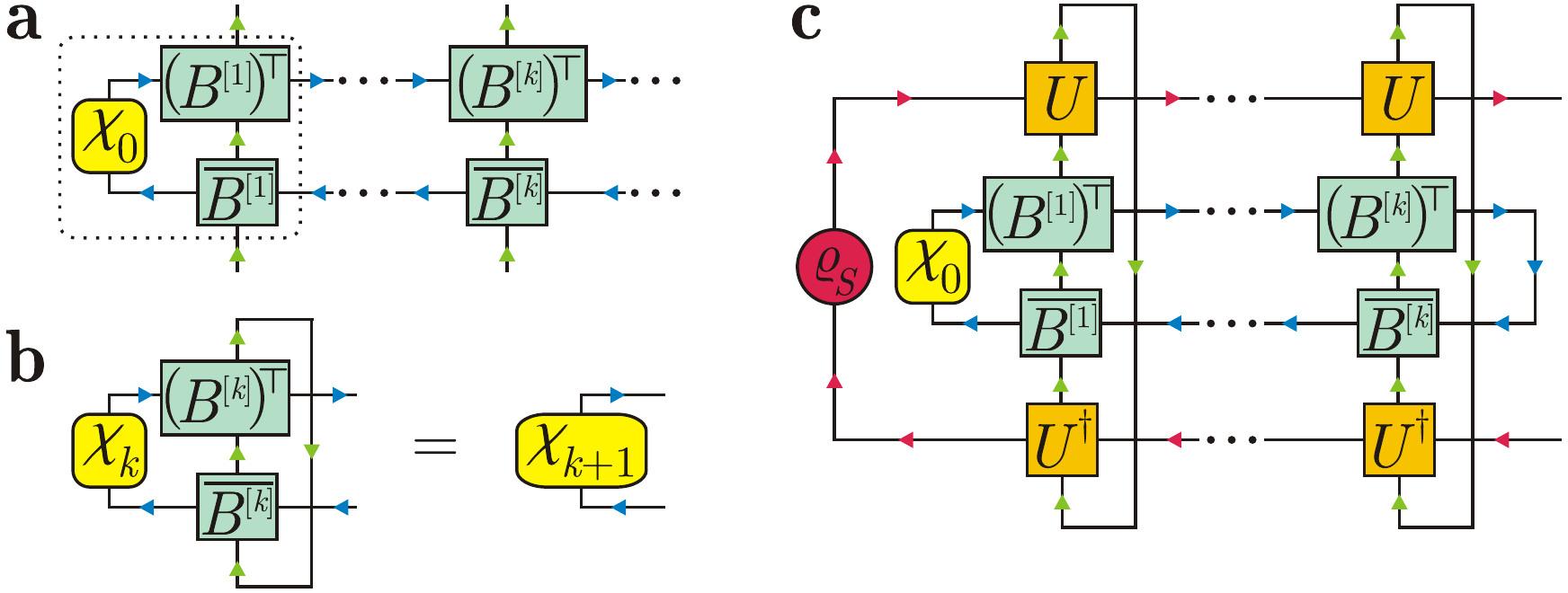}
\caption{(\textbf{a}) Matrix product density operator
representation for a generally correlated state of ancillas, where
we have formally added an auxiliary density matrix $\chi_0$ for
the bond degrees of freedom to redefine the first MPDO tensor
(dotted region). (\textbf{b}) Recurrence relation on density
operators for bond degrees of freedom in the interaction-free
evolution. (\textbf{c}) Tensor diagram for the system density
operator $\varrho_S(k\tau)$ after $k$ collisions with correlated
ancillas.} \label{figure7}
\end{figure}

Let the intial state $\varrho_{1 \ldots n}$ be a right-canonical
MPDO for $n$ ancillas shown in Figure~\ref{figure7}a. Here, we
have added a formal density operator $\chi_0$ (i.e., a positive
semidefinite operator with unit trace) for the bond degrees of
freedom (blue arrows in Figure~\ref{figure7}a). In the
conventional MPDO notation $\chi_0$ is the trivial $1 \times 1$
identity matrix for dummy indices; however, in our construction it
can be an arbitrary density matrix such that the tensor
contraction is well defined. Note that we changed the direction of
arrows in the upper line in Figure~\ref{figure7}a. This implies
transposition of matrices $B_b^{[k],i_k}$ with respect to
horizontal virtual indices.

Figure~\ref{figure7}b illustrates that the tensors $\{B^{[k]}\}_k$
define a ``free evolution'' for the bond degrees of freedom if
ancillas do not interact with the system, namely, the matrix
\begin{equation} \label{chi-recurrence}
\chi_{k+1} = \sum_{i_k,b} (B^{[k],i_k}_{b})^{\top} \chi_k
\overline{B^{[k],i_k}_{b}}
\end{equation}

\noindent is a valid density matrix (i.e., $\chi_{k+1}^{\dag} =
\chi_{k+1} \geq 0$ and ${\rm tr}[\chi_k] = 1$) provided $\chi_k$
is a density matrix too. This follows from the right normalization
condition~\eqref{right-normalization-mpdo}.

Figure~\ref{figure7}c depicts the system density operator
$\varrho_S(k\tau)$ after $k$ collisions. The partial trace over
ancillas $k+1, \ldots, n$ corresponds to a vertical connecting
line for the bond degrees of freedom (blue arrows in
Figure~\ref{figure7}c). Partial trace over ancillas $1, \ldots, k$
corresponds to vertical connecting lines for the ancillary degrees
of freedom (green arrows in Figure~\ref{figure7}c).
Ref.~\cite{filippov-2022} discusses the natural Markovian
embedding for the system dynamics that follows from the diagram in
Figure~\ref{figure7}c. In our case, we have
\begin{equation} \label{system-dynamics}
\varrho_S(k\tau) = {\rm tr}_{\rm bond} \left[ {\cal E}^{[k]} \circ
\ldots \circ {\cal E}^{[1]} [\varrho_S \otimes \chi_0] \right],
\end{equation}

\noindent where $\circ$ denotes a map concatenation, each map
${\cal E}^{[m]}$ is completely positive and trace preserving
because it adopts the diagonal sum representation
\begin{equation} \label{E-map}
{\cal E}^{[m]}[R] = \sum_{j_m b} K_{j_m b} \, R \, K_{j_m
b}^{\dag}, \quad K_{j_m b} = \sum_{i_m} \Big( (I_S \otimes
\bra{j_m}) U (I_S \otimes \ket{i_m}) \Big) \otimes
(B^{[m],i_m}_{b})^{\top}.
\end{equation}

\noindent The trace preserving property $\sum_{j_m b} K_{j_m
b}^{\dag} K_{j_m b} = I$ follows from the right normalization
condition~\eqref{right-normalization-mpdo} and unitarity of $U$.

The tensor diagram in Figure~\ref{figure7}c is a particular form
of the process tensor --- a recently developed approach to an
operational description of non-Markovian quantum
dynamics~\cite{taranto-2019,pollock-pra-2018,pollock-prl-2018,white-2020,taranto-2020}.
Complexity of the non-Markovian dynamics simulation depends on the
dimension of the effective reservoir in the Markovian
embedding~\cite{luchnikov-2019,luchnikov-2020}: the less the
dimension of the Markovian embedding the simpler the simulation.
In our model, the role of the effective reservoir is played by the
bond degrees of freedom that specify correlations among the
ancillas.

Emergence of non-Markovian dynamics in the case of correlated
ancillas was demonstrated in Ref.~\cite{rybar-2012}, where an
exemplary indecomposable qubit channel was realized as a result of
qubit's collisional interactions with many qutrit ancillas in the
GHZ state. The analytical treatment in Ref.~\cite{rybar-2012} was
only possible due to a peculiar controlled-unitary qubit-ancilla
interaction. Were the qubit-ancilla interaction different from the
controlled-unitary type, the methods of Ref.~\cite{rybar-2012}
would not provide any analytical expression for the qubit system
dynamics (nor would it be possible to study its non-Markovianity).
As we show in the example below, the developed tensor network
formalism enables us to resolve that difficulty and analytically
derive the qubit dynamics even for non-controlled-unitary
collisions. Since any environment state adopts an MPDO form, our
results generalize those of Ref.~\cite{filippov-2017}, where
non-Markovian qubit dynamics is induced by a specific correlated
environment $\varrho_{1\ldots n} = \oplus_{m} p_m \otimes_{k=1}^n
\varrho_k^{(m)}$ or $\varrho_{1\ldots n} = \otimes_{m=x,y,z}
\left( \frac{1}{2} \otimes_{k \in \{k_m\}} \ket{i_k} \bra{i_k} +
\frac{1}{2} \otimes_{k \in \{k_m\}} \ket{\overline{i_k}}
\bra{\overline{i_k}}\right)$, where either $i_k = 0$ and
$\overline{i_k} = 1$, or $i_k = 1$ and $\overline{i_k} = 0$;
$\{k_x\}$, $\{k_y\}$, $\{k_z\}$ are nonintersecting subsequences
of collision numbers. The latter environment reproduces an
arbitrary Pauli dynamical map~\cite{filippov-2017}.

\begin{figure}[b]
\centering
\includegraphics[width=16cm]{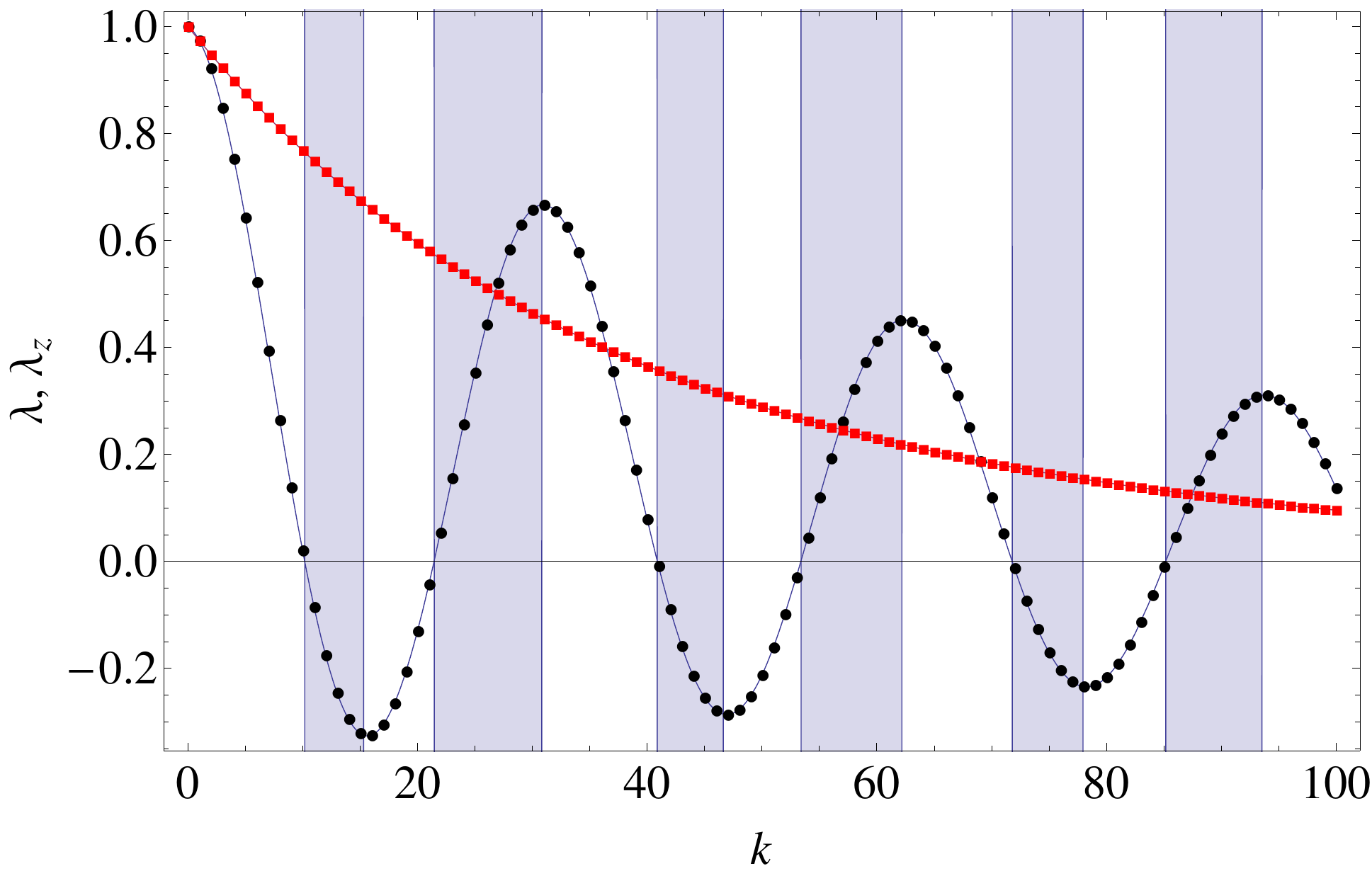}
\caption{Dephasing coefficient $\lambda$ (black circles) and
amplitude relaxation coefficient $\lambda_z$ (red squares) vs
number of collisions $k$ in qubit
dynamics~\eqref{unital-qubit-example} with $g\tau = 0.2$ emerging
in the collision model with the correlated GHZ state of qutrit
ancillas. Colored regions correspond to essential non-Markovianity
(positive indivisibility) of the qubit dynamics.} \label{figure8}
\end{figure}

\begin{example}
Consider the GHZ state of 3-dimensional ancillas $\varrho_{1
\ldots n} = \ket{{\rm GHZ}}\bra{{\rm GHZ}}$, $\ket{{\rm GHZ}} =
\frac{1}{\sqrt{3}} \sum_{j=1,2,3} \ket{j}^{\otimes n}$. The tensor
network representation for this pure state reads

\begin{equation} \label{chi-B-GHZ}
\chi_0 = \frac{1}{3} \left(%
\begin{array}{ccc}
  1 & 1 & 1 \\
  1 & 1 & 1 \\
  1 & 1 & 1 \\
\end{array}%
\right), \quad B_1^{[k],1} = \left(%
\begin{array}{ccc}
  1 & 0 & 0 \\
  0 & 0 & 0 \\
  0 & 0 & 0 \\
\end{array}%
\right), \quad B_1^{[k],2} = \left(%
\begin{array}{ccc}
  0 & 0 & 0 \\
  0 & 1 & 0 \\
  0 & 0 & 0 \\
\end{array}%
\right), \quad B_1^{[k],3} = \left(%
\begin{array}{ccc}
  0 & 0 & 0 \\
  0 & 0 & 0 \\
  0 & 0 & 1 \\
\end{array}%
\right).
\end{equation}

\noindent The qubit system is initially in the state $\varrho_S
\equiv \varrho_S(0)$. Each qubit-qutrit collision is described by
the unitary operator
\begin{equation} \label{U-sigma-J}
U = \exp \left[ - i \frac{g \tau}{2} \sum_{j=1,2,3} \sigma_j
\otimes J_j \right],
\end{equation}

\noindent where $(\sigma_1,\sigma_2,\sigma_3) \equiv
(\sigma_x,\sigma_y,\sigma_z)$ is the conventional set of Pauli
operators, $(J_1,J_2,J_3) \equiv (J_x,J_y,J_z)$ is a set of SU(2)
generators for a qutrit (spin-1 particle). In the conventional
orthonormal basis $(\ket{1},\ket{2},\ket{3})$ the corresponding
matrices are
\begin{equation} \label{J-formula}
J_x = \frac{1}{\sqrt{2}} \left(%
\begin{array}{ccc}
  0 & 1 & 0 \\
  1 & 0 & 1 \\
  0 & 1 & 0 \\
\end{array}%
\right), \quad J_y = \frac{1}{\sqrt{2}} \left(%
\begin{array}{ccc}
  0 & -i & 0 \\
  i & 0 & -i \\
  0 & i & 0 \\
\end{array}%
\right), \quad J_z = \left(%
\begin{array}{ccc}
  1 & 0 & 0 \\
  0 & 0 & 0 \\
  0 & 0 & -1 \\
\end{array}%
\right).
\end{equation}

\noindent Substituting Eqs.~\eqref{chi-B-GHZ}
and~\eqref{U-sigma-J} into Eq.~\eqref{E-map}, we get the map
${\cal E}^{[m]} \equiv {\cal E}$ that does not depend on collision
number $m$. Then Eq.~\eqref{system-dynamics} results in the
following qubit system density operator after $k$ collisions:
\begin{equation} \label{unital-qubit-example}
\varrho_S(k\tau) = \frac{1}{2} \left\{ I + \lambda(k) {\rm
tr}[\varrho_S(0) \sigma_x] \sigma_x + \lambda(k) {\rm
tr}[\varrho_S(0) \sigma_y] \sigma_y + \lambda_z(k) {\rm
tr}[\varrho_S(0) \sigma_z] \sigma_z \right\},
\end{equation}

\noindent where $\lambda(k)$ is a scaling coefficient for the $x$
and $y$ components of the qubit Bloch vector, $\lambda_z(k)$ is a
scaling coefficient for the $z$ component of the qubit Bloch
vector. The center of the Bloch ball is a steady point under such
dynamics. The explicit formulas for the scaling coefficients are
\begin{eqnarray}
\lambda(k) &=& \frac{3^k[1+2\exp(3ig\tau/2)]^k + 3^k[1+2\exp(-3ig\tau/2)]^k + [5+4\cos(3g\tau/2)]^k}{3^{2k+1}} \, , \quad \\
\lambda_z(k) &=& \frac{[1+8 \cos(3g\tau/2)]^k + 2
[5+4\cos(3g\tau/2)]^k}{3^{2k+1}} \, .
\end{eqnarray}

\noindent We portray the typical behaviour of $\lambda(k)$ and
$\lambda_z(k)$ in Figure~\ref{figure8}. Whenever $|\lambda(k)|$
increases with the increase of $k$, we observe a positive
indivisible dynamics, which is often treated as an indication of
essential non-Markovianity~\cite{chruscinski-maniscalco-2014}. We
refer the interested reader to Ref.~\cite{fgl-2020} for a full
analysis of divisibility properties under the phase covariant
qubit dynamics
--- the class of quantum dynamical maps comprising the
map~\eqref{unital-qubit-example}. \hfill$\triangle$\end{example}

\section{Master equation}

Eq.~\eqref{system-dynamics} defines the discrete dynamical map
$\Upsilon_{k\tau}$ that transforms the initial system density
operator $\varrho_S(0) \equiv \varrho_S$ to the system density
operator $\varrho_S(k\tau)$ after $k$ collisions, i.e.,
$\Upsilon_{k\tau}[\varrho_S(0)] = \varrho_S(k\tau)$. If $\tau
\rightarrow 0$, then we interpret $k\tau$ as a continuous time
$t$. A time-local master equation $\frac{d\varrho_S(t)}{dt} = L_t
[\varrho_S(t)]$ can be derived if $\Upsilon_t$ is invertible,
namely, $L_t = \frac{d \Upsilon_t}{dt} \circ \Upsilon_t^{-1}$.
Although such a master equation correctly describes the system
evolution, it conceals the major role of correlations among
ancillas. To reveal the physics of how these correlations affect
the system dynamics, we resort to the conventional projection
operator techniques~\cite{breuer-2002} and derive the
Nakajima-Zwanzig memory-kernel
equation~\cite{nakajima-1958,zwanzig-1960} for our model.

Let $\{\chi_k\}_{k}$ be a collection of the density operators for
the bond degrees of freedom generated by
Eq.~\eqref{chi-recurrence}. For each $k$ define the following map
$P_k$ acting on both the system and the bond degrees of freedom:
\begin{equation} \label{projector-P}
P_k [R] = {\rm tr}_{\text{bond}} [R] \otimes \chi_k.
\end{equation}

\begin{figure}[b]
\centering

\includegraphics[width=18cm]{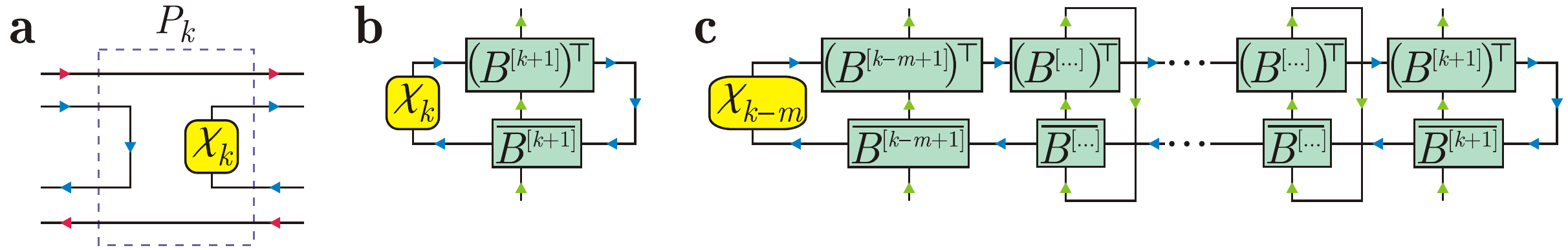}
  \caption{(\textbf{a}) Projection $P_k$.
(\textbf{b}) Reduced density operator $\varrho_{k+1}$ for the
$(k+1)$-th ancilla in the initial correlated state of ancillas.
(\textbf{c}) Reduced density operator $\varrho_{k-m+1,k+1}$ for
the $(k-m+1)$-th ancilla and $(k+1)$-th ancilla in the initial
correlated state of ancillas.} \label{figure9}
\end{figure}

\noindent Since ${\rm tr}[\chi_k] = 1$, we have $P_k^2 = P_k$ so
$P_k$ is a projection. A pictorial representation of the
projection $P_k$ is given in Figure~\ref{figure9}a, which shows
that $P_k$ breaks the left-right correlations between bunches of
ancillas $(1,\ldots,k)$ and $(k+1,\ldots,n)$. Clearly,
\begin{equation} \label{P-to-rho}
P_k \left[ {\cal E}^{[k]} \circ \ldots \circ {\cal E}^{[1]}
[\varrho_S(0) \otimes \chi_0] \right] = \varrho_S(k\tau) \otimes
\chi_k.
\end{equation}

A complementary projection $Q_k$ is defined through
\begin{equation}
Q_k = {\rm Id}_{S+\text{bond}} - P_k,
\end{equation}

\noindent where ${\rm Id}$ is the identity transformation for the
system and bond degrees of freedom. $Q_k$ is a projection too
because $Q_k^2=Q_k$. We can also rewrite $Q_k$ in the form
\begin{equation}
Q_k = {\rm Id}_S \otimes Q_{\text{bond} \# k},
\end{equation}

\noindent where $Q_{\text{bond} \# k}$ is a projection for the
$k$-th bond degrees of freedom that acts on an operator $F$ for
the bond degrees of freedom as follows:
\begin{equation} \label{Q-bond}
Q_{\text{bond} \# k} [F] = F - {\rm tr}[F] \chi_k.
\end{equation}

\noindent Since $Q_{\text{bond} \# 0} [\chi_0] = 0$, we readily
get
\begin{equation} \label{Q-initial}
Q_0 [ \varrho_S(0) \otimes \chi_0 ] = 0.
\end{equation}

To simplify the notation, let us introduce the system-bond density
operator after $k$-th collision, $R(k\tau) = {\cal E}^{[k]} \circ
\ldots \circ {\cal E}^{[1]} [\varrho_S(0) \otimes \chi_0]$. Then
$[ R(k\tau)] = {\cal E}^{[k]} \left[ R\big((k-1)\tau \big)
\right]$ for all $k$. Applying $Q_k$ to the both sides of the
latter equation, we get
\begin{equation} \label{Q-recurrent}
Q_k[ R(k\tau)] = Q_k \circ {\cal E}^{[k]} \circ (P_{k-1}+Q_{k-1})
\left[ R\big((k-1)\tau \big) \right].
\end{equation}

\noindent The recurrent equation~\eqref{Q-recurrent} with the
initial condition~\eqref{Q-initial} has the following formal
solution:
\begin{equation} \label{Q-solution}
Q_k[ R(k\tau)] = \sum_{m=1}^k Q_k \circ {\cal E}^{[k]} \circ
\ldots \circ Q_{k-m+1} \circ {\cal E}^{[k-m+1]} \circ P_{k-m}
\left[ R\big((k-m)\tau \big) \right].
\end{equation}

\noindent If we apply $P_{k+1}$ to the both sides of equation $[
R((k+1)\tau \big)] = {\cal E}^{[k+1]} \left[ R(k\tau) \right]$, we
obtain

\begin{eqnarray}
&& P_{k+1}[ R \big( (k+1)\tau \big) ] = P_{k+1} \circ {\cal
E}^{[k+1]} \circ (P_k + Q_k) \left[ R(k\tau) \right] = P_{k+1}
\circ {\cal E}^{[k+1]} \circ P_k \left[ R(k\tau) \right] + P_{k+1}
\circ {\cal E}^{[k+1]} Q_k \left[ R(k\tau) \right] \nonumber\\
&& = P_{k+1} \circ {\cal E}^{[k+1]} \circ P_k \left[ R(k\tau)
\right] + \sum_{m=1}^k P_{k+1} \circ  {\cal E}^{[k+1]} \circ Q_k
\circ {\cal E}^{[k]} \circ \ldots \circ Q_{k-m+1} \circ {\cal
E}^{[k-m+1]} \circ P_{k-m} \left[ R\big((k-m)\tau \big) \right].
\qquad \label{P-recurrent}
\end{eqnarray}

\noindent Recalling the relation $P_k \left[ R(k\tau) \right] =
\varrho_S(k\tau) \otimes \chi_k$ and taking partial trace over the
bond indices in Eq.~\eqref{P-recurrent}, we get

\begin{equation} \label{rho-recurrent}
\varrho_S \big( (k+1)\tau \big) = {\rm tr}_{\text{bond}} \circ
{\cal E}^{[k+1]} [ \varrho_S(k\tau) \otimes \chi_k] + \sum_{m=1}^k
{\rm tr}_{\text{bond}} \circ {\cal E}^{[k+1]} \circ Q_k \circ
{\cal E}^{[k]} \circ \ldots \circ Q_{k-m+1} \circ {\cal
E}^{[k-m+1]} [ \varrho_S \big((k-m)\tau \big) \otimes \chi_{k-m} ]
.
\end{equation}

\noindent Subtracting $\varrho_S(k\tau)$ from both sides of
Eq.~\eqref{rho-recurrent} and dividing the result by the collision
time $\tau$, we get a discrete-time version of the celebrated
Nakajima-Zwanzig master equation, namely,
\begin{equation} \label{nz-discrete}
\frac{\varrho_S \big( (k+1)\tau \big) - \varrho_S(k\tau)}{\tau} =
\sum_{m=0}^{k} {\cal K}_{km} [ \varrho_S \big((k-m)\tau \big) ],
\end{equation}

\noindent where the memory kernel ${\cal K}_{km}$ relates the
density matrix increment [in between the times $k\tau$ and
$(k+1)\tau$] with the past density operator at time $(k-m)\tau$.
If $m=0$, then we have a time-local term ${\cal K}_{k0}$ giving
the density operator increment caused by the latest collision
(among those that have already happened):
\begin{equation}
{\cal K}_{k0} [\varrho_S] = \frac{{\rm tr}_{k+1} [U \varrho_S
\otimes \varrho_{k+1} U^{\dag}] - \varrho_S}{\tau}, \label{K-k0}
\end{equation}

\noindent with $\varrho_{k+1}$ being a reduced density operator
for $(k+1)$-th ancilla in the initial state, see
Figure~\ref{figure9}b. If $m \geq 1$, then ${\cal K}_{km}$
describes a nontrivial effect of preceding collisions on the
system evolution and reads
\begin{equation}
{\cal K}_{km} [\varrho_S] = \frac{1}{\tau} {\rm tr}_{\text{bond}}
\circ {\cal E}^{[k+1]} \circ Q_k \circ {\cal E}^{[k]} \circ \ldots
\circ Q_{k-m+1} \circ {\cal E}^{[k-m+1]} [\varrho_S \otimes
\chi_{k-m}]. \label{K-km}
\end{equation}

\noindent If there were no correlations in the environment, then
${\cal K}_{k0}$ would be the only contribution to the kernel
because ${\cal K}_{km}$ would vanish for all $m \geq 1$. Indeed,
the MPDO rank equals 1 for a factorized environment state, so
${\rm dim} {\cal H}_{\text{bond} \# k} = 1$ for all $k$, each
$\chi_k$ is unambiguously defined because $\chi_k$ the trivial $1
\times 1$ identity matrix in this case, and $Q_{\text{bond} \#
k}[F] = 0$ for any $1\times 1$ matrix $F$. If the environment is
correlated, then the memory contribution ${\cal K}_{km}[\varrho_S]
\neq 0$ in general.

\begin{example}
Consider the GHZ state of 3-dimensional ancillas $\varrho_{1
\ldots n} = \ket{{\rm GHZ}}\bra{{\rm GHZ}}$, $\ket{{\rm GHZ}} =
\frac{1}{\sqrt{3}} \sum_{j=1,2,3} \ket{j}^{\otimes n}$ and the
controlled unitary system-ancilla interaction $U = \sum_{j=1,2,3}
e^{-i g \tau \sigma_j} \otimes \ket{j}\bra{j}$, where $g\tau$
quantifies the dimensionless system-ancilla interaction strength,
$(\sigma_1,\sigma_2,\sigma_3) \equiv (\sigma_x,\sigma_y,\sigma_z)$
is the conventional set of Pauli operators. This is a scenario
considered also Ref.~\cite{rybar-2012}. A direct calculation
yields

\begin{equation}
{\cal K}_{km} [\varrho_S] = \frac{1}{3\tau}
\sum_{i_{k+1},\ldots,i_{k-m+1} = 1}^3 \left[ \prod_{l=k-m+1}^{k}
\left( \delta_{i_l,i_{l+1}}  - \frac{1}{3} \right) \right]
e^{-ig\tau \sigma_{i_{k+1}}} \cdots e^{-ig\tau \sigma_{i_{k-m+1}}}
\varrho_S e^{ig\tau \sigma_{i_{k-m+1}}} \cdots e^{ig\tau
\sigma_{i_{k+1}}}.
\end{equation}

\noindent The memory kernel ${\cal K}_{km}$ does not decay with
the increase of $m$ due to the infinite correlation length in the
GHZ state. In view of this, even if the interaction strength
$g\tau \ll 1$, one cannot truncate a series expansion for ${\cal
K}_{km}$ with respect to a small parameter $g\tau$. Instead, all
orders of $g\tau$ are significant for reproducing the system
dynamics. \hfill$\triangle$\end{example}

If the correlation length is finite, then it is possible to derive
a continuous-time master equation in the appropriate limit for
$\tau$ and $g$. This is discussed in what follows.

\section{Effect of two-point correlations}

An important simplification comes from a series expansion for
$U_{Sm}$ with respect to the interaction strength $g \tau$ between
the system and an individual environment particle. Let $g \hbar
H_m$ be the system-particle interaction Hamiltonian during the
$m$-th collision, where $\hbar$ is the reduced Planck constant,
$g$ has the physical dimension of frequency, and $H_m$ is a
dimensionless Hermitian operator with the operator norm $\|H_m\|
\leq 1$. Then the elementary unitary interaction in the $m$-th
collision is $U = \exp(-i g \tau H_m)$. The map ${\cal E}^{[m]}$
in Eq.~\eqref{E-map} has a contribution of both $U$ and
$U^{\dag}$, so we have

\begin{eqnarray}
&& {\cal E}^{[k]} = \sum_{i_k,i'_k} \Phi^{[k]}_{i_k i'_k} \otimes
\Lambda^{[k]}_{i_k i'_k}, \\
&& \Phi^{[k]}_{i_k i'_k}[\varrho_S] \equiv {\rm tr}_k \left[ U \,
\varrho_S \otimes \ket{i_k}\bra{i'_k} \, U \right] \nonumber\\
&& \qquad = \delta_{i_k i'_k} \varrho_S - i g \tau \Big[
\bra{i'_k} H_k \ket{i_k}, \varrho_S \Big] + g^2 \tau^2 \bigg(
\sum_{j_k = 1}^d \bra{j_k} H_k \ket{i_k} \varrho_S \bra{i'_k} H_k
\ket{j_k} - \frac{1}{2} \Big\{ \bra{i'_k} H_k^2 \ket{i_k} ,
\varrho_S \Big\} \bigg)  + o(g^2 \tau^2) \nonumber\\
&& \qquad \equiv  \Phi^{[k],{(0)}}_{i_k i'_k}[\varrho_S] + g\tau
\Phi^{[k],{(1)}}_{i_k i'_k}[\varrho_S] + g^2 \tau^2
\Phi^{[k],{(2)}}_{i_k i'_k}[\varrho_S] + o(g^2 \tau^2),
\label{Phi-expansion} \\
&& \Lambda^{[k]}_{i_k i'_k} [ \bullet ] = \sum_b
(B^{[k],i'_k}_b)^{\top} \bullet \overline{B^{[k],i_k}_b},
\end{eqnarray}

\noindent where $[\bullet,\bullet]$ and $\{\bullet,\bullet\}$
denote the commutator and the anticommutator, respectively.
Substituting Eq.~\eqref{Phi-expansion} into Eq.~\eqref{K-km}, we
keep track of the leading terms in the memory kernel, namely,
\begin{equation} \label{kernel-series}
{\cal K}_{km} = \frac{1}{\tau} {\cal K}^{(0)}_{km} + g {\cal
K}^{(1)}_{km} + g^2 \tau {\cal K}^{(2)}_{km} + o(g^2\tau).
\end{equation}

The term ${\cal K}^{(0)}_{km}$ vanishes because
\begin{equation}
Q_{\text{bond} \# (k-m+1)} \circ \sum_{i_{k-m+1}}
\Lambda^{[k-m+1]}_{i_{k-m+1} i_{k-m+1}} [\chi_{k-m}] =
Q_{\text{bond} \# (k-m+1)} [\chi_{k-m+1}] = 0, \label{kernel-zero}
\end{equation}

\noindent see Eq.~\eqref{Q-bond}. Physically, the $0$-th order of
$\Phi^{[k]}_{i_k,i'_k}$ involves no system-environment interaction
and, consequently, no contribution to the memory kernel.

To calculate the term ${\cal K}^{(1)}_{km}$ we should fix
$\Phi^{[l]}_{i_l,i'_l} = \delta_{i_l i'_l} {\rm Id}_S \equiv
\Phi^{[l],(0)}_{i_l,i'_l}$ for all but one of $l \in
(k-m+1,\ldots,k+1)$. If $\Phi^{[k-m+1]}_{i_{k-m+1},i'_{k-m+1}} =
\delta_{i_{k-m+1} i'_{k-m+1}} {\rm Id}_S$, then we have a zero
contribution to ${\cal K}^{(1)}_{km}$ because of
Eq.~\eqref{kernel-zero}. Suppose
$\Phi^{[k-m+1]}_{i_{k-m+1},i'_{k-m+1}} \neq \delta_{i_{k-m+1}
i'_{k-m+1}} {\rm Id}_S$, then $\Phi^{[k+1]}_{i_{k+1},i'_{k+1}} =
\delta_{i_{k+1} i'_{k+1}} {\rm Id}_S$ and
\begin{equation}
{\rm tr} \Big[ \sum_{i_{k+1}} \Lambda^{[k+1]}_{i_{k+1} i_{k+1}}
\circ Q_{\text{bond} \# k} [\bullet] \Big] = {\rm tr} \Big[
Q_{\text{bond} \# k} [\bullet] \Big] = 0, \label{kernel-zero-2}
\end{equation}

\noindent because $\sum_{i_{k+1}} \Lambda^{[k+1]}_{i_{k+1}
i_{k+1}}$ is a trace preserving map due to the right-normalization
condition, whereas $Q_{\text{bond} \# k}$ nullifies the trace of
any operator, see Eq.~\eqref{Q-bond}. Therefore, the term ${\cal
K}^{(1)}_{km}$ vanishes too.

Similar considerations for the term ${\cal K}^{(2)}_{km}$ lead to
a conclusion that ${\cal K}^{(2)}_{km}$ may be nonzero only if we
fix $\Phi^{[l]}_{i_l,i'_l} = g\tau \Phi^{[l],(1)}_{i_l,i'_l}$
$\equiv - i g \tau \Big[ \bra{i'_l} H_l \ket{i_l},\varrho_S \Big]
$ for $l=k+1$ and $l=k-m+1$, whereas for all $l=k-m+2,\ldots,k$ we
fix $\Phi^{[l]}_{i_l,i'_l} = \Phi^{[l],(0)}_{i_l,i'_l} \equiv
\delta_{i_l i'_l} {\rm Id}_S$. This results in

\begin{eqnarray}
&& {\cal K}_{km}^{(2)} [\varrho_S] = \sum_{i_{k+1},i'_{k+1},
i_{k-m+1},i'_{k-m+1}} C^{(2)}_{ i_{k-m+1} i'_{k-m+1} i_{k+1}
i'_{k+1} }  \Phi^{[k+1],{(1)}}_{i_{k+1} i'_{k+1}} \circ
\Phi^{[k-m+1],{(1)}}_{i_{k-m+1} i'_{k-m+1}}[\varrho_S]
\nonumber\\
&& = - \sum_{i_{k+1},i'_{k+1}, i_{k-m+1},i'_{k-m+1}} C^{(2)}_{
i_{k-m+1} i'_{k-m+1} i_{k+1} i'_{k+1} } \bigg[ \bra{i'_{k+1}}
H_{k+1} \ket{i_{k+1}} , \Big[ \bra{i'_{k-m+1}} H_{k-m+1}
\ket{i_{k-m+1}} , \varrho_S \Big] \bigg], \label{K-km-2nd-order}
\end{eqnarray}

\noindent where the coefficient $C^{(2)}_{i_{k-m+1} i'_{k-m+1}
i_{k+1} i'_{k+1}}$ reads

\begin{eqnarray} \label{C-complex}
&& C^{(2)}_{i_{k-m+1} i'_{k-m+1} i_{k+1} i'_{k+1}} \nonumber\\
&& = {\rm tr} \Big[ \Lambda^{[k+1]}_{i_{k+1} i'_{k+1}} \circ
Q_{\text{bond} \# k} \circ \sum_{i_k} \Lambda^{[k]}_{i_k i_k}
\circ \ldots \circ Q_{\text{bond} \# (k-m+2)} \circ
\sum_{i_{k-m+2}} \Lambda^{[k-m+2]}_{i_{k-m+2} i_{k-m+2}} \circ
Q_{\text{bond} \# (k-m+1)} \circ \Lambda^{[k-m+1]}_{i_{k-m+1}
i'_{k-m+1}} [\chi_{k-m}] \Big]. \nonumber\\
\end{eqnarray}

\noindent Recalling the definition~\eqref{Q-bond}, we get

\begin{equation} \label{C-interm}
Q_{\text{bond} \# k} \circ \sum_{i_k} \Lambda^{[k]}_{i_k i_k}
\circ \ldots \circ Q_{\text{bond} \# (k-m+2)} \circ
\sum_{i_{k-m+2}} \Lambda^{[k-m+2]}_{i_{k-m+2} i_{k-m+2}} \circ
Q_{\text{bond} \# (k-m+1)} = \sum_{i_k,\ldots,i_{k-m+2}}
\Lambda^{[k]}_{i_k i_k} \circ \ldots \circ
\Lambda^{[k-m+2]}_{i_{k-m+2} i_{k-m+2}} - P_k.
\end{equation}

\noindent Tensor representation in Figure~\ref{figure9}c justifies
that
\begin{eqnarray} \label{C-1}
{\rm tr} \Big[ \Lambda^{[k+1]}_{i_{k+1} i'_{k+1}} \circ
\sum_{i_k,\ldots,i_{k-m+2}} \Lambda^{[k]}_{i_k i_k} \circ \ldots
\circ \Lambda^{[k-m+2]}_{i_{k-m+2} i_{k-m+2}} \circ
\Lambda^{[k-m+1]}_{i_{k-m+1} i'_{k-m+1}} [\chi_{k-m}] \Big] =
\varrho_{k-m+1,k+1},
\end{eqnarray}

\noindent i.e., we get the reduced density operator for the
$(k-m+1)$-th ancilla and $(k+1)$-th ancilla in the initial
correlated state of ancillas. Similarly,
\begin{eqnarray} \label{C-2}
{\rm tr} \Big[ \Lambda^{[k+1]}_{i_{k+1} i'_{k+1}} \circ P_k \circ
\Lambda^{[k-m+1]}_{i_{k-m+1} i'_{k-m+1}} [\chi_{k-m}] \Big] =
\varrho_{k-m+1} \otimes \varrho_{k+1},
\end{eqnarray}

\noindent i.e., we get a tensor product of individual reduced
density operators for the $(k-m+1)$-th ancilla and $(k+1)$-th
ancilla in the initial correlated state of ancillas.

Combining \eqref{C-complex}, \eqref{C-interm}, \eqref{C-1}, and
\eqref{C-2}, we obtain a surprisingly simple though exact result,
namely,

\begin{equation}
C^{(2)}_{i_{k-m+1} i'_{k-m+1} i_{k+1} i'_{k+1}} = \bra{i_{k-m+1}
i_{k+1}} \varrho_{k-m+1,k+1} \ket{i'_{k-m+1} i'_{k+1}} -
\bra{i_{k-m+1}} \varrho_{k-m+1} \ket{i'_{k-m+1}} \bra{i_{k+1}}
\varrho_{k+1} \ket{i'_{k+1}}. \label{C-correlator}
\end{equation}

\noindent Introducing the environment two-point correlation
function for operators $O$ and $O'$ by a conventional formula
\begin{eqnarray}
{\cal C}(O,O') &=& \braket{OO'}_{\rm anc} - \braket{O}_{\rm anc}
\braket{O'}_{\rm anc} \label{correlator} \\
&=& {\rm tr} \left[ O \otimes O' (\varrho_{k-m+1,k+1} -
\varrho_{k-m+1} \otimes \varrho_{k+1})\right],
\end{eqnarray}

\noindent we readily see that $C^{(2)}_{i_{k-m+1} i'_{k-m+1}
i_{k+1} i'_{k+1}} = {\cal C}(O,O')$, where
$O=\ket{i'_{k-m+1}}\bra{i_{k-m+1}}$ and
$O'=\ket{i'_{k+1}}\bra{i_{k+1}}$. Combining all the findings of
this section, we get

\begin{eqnarray}
{\cal K}_{km} [\varrho_S] &=& - g^2 \tau \sum_{i_{k+1},i'_{k+1},
i_{k-m+1},i'_{k-m+1}} {\cal C}(\ket{i'_{k-m+1}}\bra{i_{k-m+1}} ,
\ket{i'_{k+1}}\bra{i_{k+1}}) \nonumber\\
&& \qquad \qquad \times \bigg[ \bra{i'_{k+1}} H_{k+1}
\ket{i_{k+1}} , \Big[ \bra{i'_{k-m+1}} H_{k-m+1} \ket{i_{k-m+1}} ,
\varrho_S \Big] \bigg] + o(g^2\tau) \nonumber\\
&=& g^2 \tau \Big[ \braket{H_{k+1}}_{\rm anc} , \big[
\braket{H_{k-m+1}}_{\rm anc} , \varrho_S \big] \Big] - g^2 \tau
\left\langle \Big[ H_{k+1} , \big[H_{k-m+1}, \varrho_S \otimes
I_{\rm anc} \big] \Big] \right\rangle_{\rm anc} + o(g^2\tau).
\label{K-km-via-correlator}
\end{eqnarray}

\noindent Eq.~\eqref{K-km-via-correlator} provides an important
physical link between the two-point correlation function of
ancillas and the memory kernel.

\section{Stroboscopic limit}

To simplify the analysis, let us assume that the correlated
ancillas are initially in the homogeneous right-canonical MPDO,
i.e., the tensors $B^{[k]}$ coincide for all $k=1,\ldots,n$ and
MPDO is fully described by the density matrix $\chi_0$ and the
tensor $M$. In this case, all local density operators for
individual ancillas coincide $\varrho_1 = \ldots = \varrho_n$;
however, the two-ancilla density operator $\varrho_{12} \neq
\varrho_1 \otimes \varrho_2$. If ancillas are initially in such a
homogeneous state, we can expect that the kernel ${\cal K}_{km}$
depends on $m$ only and does not depend on $k$.

Suppose the collision duration $\tau$ tends to zero while the
coupling strength $g$ remains constant. Then we get the
Hamiltonian dynamics for the system $\varrho_S(t)$ in continuous
time $t = k\tau$, namely, $\frac{d\varrho_S(t)}{dt} = - i g
[\braket{H}_{\rm anc},\varrho_S(t)]$~\cite{palma-2012}. The
correlations among ancillas are irrelevant in this scenario
because $g^2 \tau \rightarrow 0$ in the considered limit.

To reveal a nonunitary system dynamics at a long timescale one
should consider a different limit $g\tau \rightarrow 0$, $g^2\tau
= {\rm const}$~\cite{giovannetti-2012,palma-2012,lorenzo-2017},
which we refer to as the (first-order) stroboscopic limit that is
also used in the analysis of dynamics induced by indirect repeated
measurements~\cite{luchnikov-2017,grimaudo-2020}. The Hamiltonian
part $- i g [\braket{H}_{\rm anc},\varrho_S(t)]$ explodes in the
master equation because $g \rightarrow \infty$; however, this
problem disappears in a proper interaction
picture~\cite{palma-2012}.

In the stroboscopic limit, one cannot simply replace
$\frac{1}{\tau} [\varrho_S \big( (k+1)\tau \big) - \varrho_S
(k\tau)]$ in the left hand side of Eq.~\eqref{nz-discrete} by
$\frac{d\varrho_S(t)}{dt}$ if the term $- i g [\braket{H}_{\rm
anc},\varrho_S(t)]$ does not vanish, because $\frac{1}{\tau}
[\varrho_S \big( (k+1)\tau \big) - \varrho_S (k\tau)] =
\frac{d\varrho_S(t)}{dt} + \frac{\tau}{2}
\frac{d^2\varrho_S(t)}{dt^2} + \ldots = \frac{d\varrho_S(t)}{dt} +
O(g^2\tau)$, and the second summand cannot be neglected. However,
if the expression $- i g [\braket{H}_{\rm anc},\varrho_S(t)]$
vanishes, then the characteristic frequency of system dynamics is
$g^2\tau$ so that $\frac{1}{\tau} [\varrho_S \big( (k+1)\tau \big)
- \varrho_S (k\tau)] = \frac{d\varrho_S(t)}{dt} + O(g^4\tau^3)$.
The second summand vanishes in the (first-order) stroboscopic
limit because $g^4 \tau^3 = (g^2\tau)^2 \tau \rightarrow 0$. The
time-local memory-kernel component should be considered in this
limit too, i.e.,
\begin{equation} \label{L-local}
\frac{{\rm tr}_{1} [U \varrho_S \otimes \varrho_{1} U^{\dag} ] -
\varrho_S}{\tau} \rightarrow L_{\rm local}[\varrho_S]
\text{~when~}g\tau \rightarrow 0, g^2\tau = {\rm const}.
\end{equation}

The higher-order contributions ${\cal K}_{km}^{(3)}, {\cal
K}_{km}^{(4)}, \ldots$ to the memory kernel~\eqref{kernel-series}
vanish only if the correlation length $l_{\rm corr}$ (in the chain
of ancillas) is finite. If this is the case, then $\|
\sum_{m=0}^{k} {\cal K}_{km}^{(N)} [ \varrho_S \big((k-m)\tau
\big) ] \| \lesssim l_{\rm corr} g^{N+1} \tau^N \rightarrow 0$ for
$N=3,4,\ldots$.

Therefore, in the first-order stroboscopic limit we get the
time-continuous master equation
\begin{eqnarray}
&& \frac{d\varrho_S(t)}{dt} = \int_0^t K(t')[\varrho_S(t-t')]dt',
\label{continuous-equation}
\\
&& K(t')[\varrho_S] = \delta(t') L_{\rm local}[\varrho_S] + g^2
\tau \lim_{\tau \rightarrow 0} \sum_{m=1}^{\infty}
\delta(t'-m\tau)
K_m[\varrho_S], \\
&& K_m[\varrho_S] = \big[ \braket{H}_{\rm anc} , [ \braket{H}_{\rm
anc} , \varrho_S ] \big] - \left\langle \big[ H_{m+1} , [H_{1},
\varrho_S \otimes I_{\rm anc} ] \big] \right\rangle_{\rm anc}.
\label{K-m-through-H}
\end{eqnarray}

\noindent The correlations are known to decay exponentially in an
MPS and an
MPDO~\cite{perez-garcia-2007,verstraete-2008,schollwock-2011,cirac-2021,orus-2014},
with the correlation length $l_{\rm corr}$ being defined by the
second largest eigenvalue of the transfer matrix $T = \sum_{i}
M^{ii}$ (in absolute values). If the correlation length is finite,
then $K_m$ represents a sum of exponentially decaying terms,
\begin{equation} \label{K-m-sum-of-L-j}
K_m[\varrho_S] = \sum_{j} (\lambda_j)^m L_{\rm
nonlocal}^{(j)}[\varrho_S],
\end{equation}

\noindent where $\{\lambda_j\}_j$ are eigenvalues of the transfer
matrix $T$ such that $|\lambda_j|< 1$ and $\{L_{\rm
nonlocal}^{(j)}\}$ are the corresponding maps.

To explicitly find the kernel $K(t')$ in
Eq.~\eqref{continuous-equation} we resort to the Laplace transform
(which is often used for the memory-kernel master
equations~\cite{chruscinski-2010,smirne-2010,filippov-2018}),
namely,
\begin{eqnarray}
K_s &=& \int_{0}^{\infty} K(t') e^{-st'} dt' = L_{\rm local} + g^2
\tau \lim_{\tau \rightarrow 0} \sum_{m=1}^{\infty}
e^{-s\tau m} K_m \nonumber\\
&=& L_{\rm local} + g^2 \tau \lim_{\tau \rightarrow 0} \sum_{j}
\left(\sum_{m=1}^{\infty} e^{-s\tau m} (\lambda_j)^m
\right) L_{\rm nonlocal}^{(j)} \nonumber\\
&=& L_{\rm local} + g^2 \tau \lim_{\tau \rightarrow 0} \sum_{j}
\frac{e^{-s\tau}\lambda_j}{1-e^{-s\tau}\lambda_j} L_{\rm
nonlocal}^{(j)} \nonumber\\
&=& L_{\rm local} + g^2 \tau \sum_{j}
\frac{\lambda_j}{1-\lambda_j} L_{\rm nonlocal}^{(j)}.
\end{eqnarray}

\noindent The result does not depend on $s$, which means the
kernel $K(t')$ becomes local in the stroboscopic limit and the
final master equation takes the form
\begin{equation} \label{master-equation-stroboscopic}
\frac{d\varrho_S(t)}{dt} = L_{\rm local}[\varrho_S(t)] + g^2 \tau
\sum_{j} \frac{\lambda_j}{1-\lambda_j} L_{\rm nonlocal}^{(j)}
[\varrho_S(t)].
\end{equation}

Physically, Eq.~\eqref{master-equation-stroboscopic} shows that if
the system quickly interacts with ancillas ($g\tau \ll 1$), then
the system ``feels'' not only the individual ancillas (which
results in the local term $L_{\rm local}$) but also a somewhat
averaged correlated state (which results in the nonlocal term $g^2
\tau \sum_{j} \frac{\lambda_j}{1-\lambda_j} L_{\rm
nonlocal}^{(j)}$). We summarize these results as follows.

\begin{proposition} \label{proposition-stroboscopic}
Let the system collisionally interact with an array of ancillas in
the homogeneous MPDO with a finite correlation length. If the
expression $- i g [\braket{H}_{\rm anc},\varrho_S(t)]$ vanishes,
then in the first-order stroboscopic limit $g\tau \rightarrow 0$,
$g^2\tau = {\rm const}$, the system dynamics is governed by the
master equation~\eqref{master-equation-stroboscopic}, where the
local and nonlocal contributions to the generator are defined by
equations~\eqref{L-local}, \eqref{K-m-through-H}, and
\eqref{K-m-sum-of-L-j}.
\end{proposition}

\begin{figure}
\centering
\includegraphics[width=12cm]{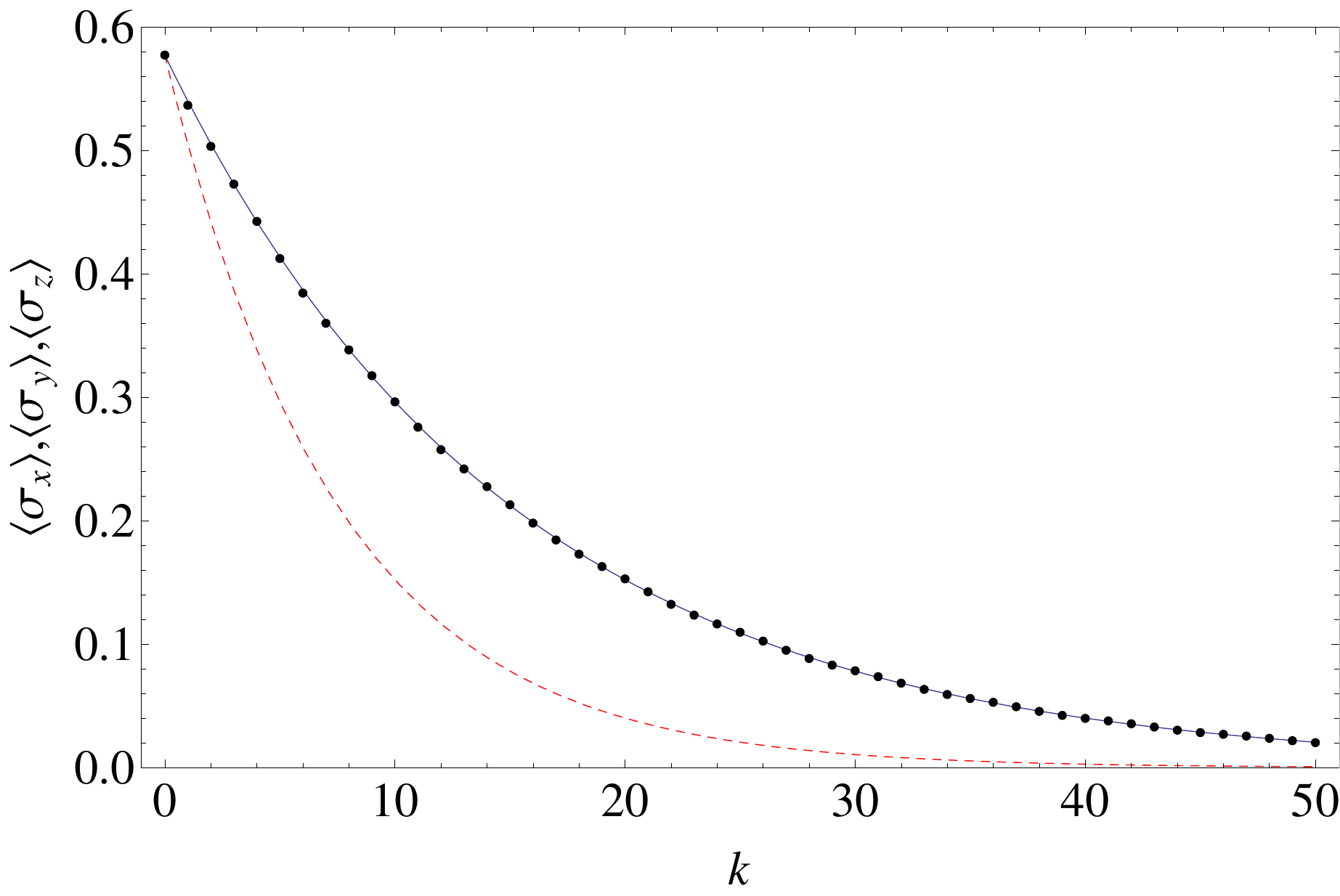}
\caption{The qubit dynamics in the collision model with correlated
ancillas in the AKLT state (Example~\ref{example-aklt-1}): Bloch
vector components
$\braket{\sigma_x},\braket{\sigma_y},\braket{\sigma_z}$ vs the
number of collisions $k$. The initial qubit state is $\varrho_S(0)
= \frac{1}{2}(I + \frac{1}{\sqrt{3}} \sum_{j=x,y,z} \sigma_j)$.
The system-ancilla Hamiltonian is given by Eq.~\eqref{H-aklt-1}.
The interaction strength $g\tau = 0.1$. Exact solution via
Eq.~\eqref{system-dynamics} is shown by dots. Solution in the
stroboscopic limit [Eq.~\eqref{equation-gksl}] corresponds to a
solid line. Disregard of correlations among ancillas leads to the
equation $\frac{d\varrho_S(t)}{dt} = L_{\rm local}[\varrho_S(t)]$,
its solution is shown by a dashed line.} \label{figure10}
\end{figure}

\begin{example} \label{example-aklt-1}
Consider an infinite chain of spin-1 particles (ancillas) in the
AKLT state~\cite{aklt-1987} that adopts the following homogenous
right canonical MPS representation with the MPS rank
$2$~\cite{schollwock-2011}:
\begin{equation}
A^{[k],1} = \left(%
\begin{array}{cc}
  0 & \sqrt{\frac{2}{3}} \\
  0 & 0 \\
\end{array}%
\right), \quad A^{[k],2} = \left(%
\begin{array}{cc}
  -\frac{1}{\sqrt{3}} & 0 \\
  0 & \frac{1}{\sqrt{3}} \\
\end{array}%
\right), \quad  A^{[k],3} = \left(%
\begin{array}{cc}
  0 & 0 \\
  -\sqrt{\frac{2}{3}} & 0 \\
\end{array}%
\right).
\end{equation}

\noindent Each individual ancilla has a reduced density operator
$\varrho_1 = \frac{1}{3}I$; however, the global state is
correlated.

At time $t=0$ the qubit system collides with one of the
intermediate ancillas, then collides with its right neighbor and
so on. Each collision lasts time $\tau$. The system-particle
interaction Hamiltonian is
\begin{equation} \label{H-aklt-1}
\hbar g H = \hbar g \sum_{j=1,2,3} \sigma_j \otimes
\ket{j}\bra{j}.
\end{equation}

\noindent Averaging over single-ancilla degrees of freedom yields
$\braket{H}_{\rm anc} = \frac{1}{3} \sum_{j=1,2,3} \sigma_j \neq
0$. To use Proposition~\ref{proposition-stroboscopic} we set
$\varrho_S(0) = \frac{1}{2}(I + \frac{1}{\sqrt{3}} \sum_{j=1,2,3}
\sigma_j)$, so that $- i g [\braket{H}_{\rm anc},\varrho_S(0)] =
0$. As we will see later, the latter commutation relation remains
valid for all times $t$, i.e., $- i g [\braket{H}_{\rm
anc},\varrho_S(t)] = 0$ and the use of
Proposition~\ref{proposition-stroboscopic} is justified.

The local term is given by formula~\eqref{L-local} and reads
\begin{equation}
L_{\rm local}[\varrho_S] = \frac{g^2\tau}{3} \sum_{j=1,2,3}
(\sigma_j \varrho_S \sigma_j - \varrho_S).
\end{equation}

To find the nonlocal term, we should take correlations into
account. Since the system interacts with a part of the infinite
spin chain, the state $\varrho_{1\ldots\infty}$ of ancillas
(spin-1 particles) is mixed and described by a right-canonical
homogeneous MPDO with

\begin{equation}
\chi_0 = \frac{1}{2} \left(%
\begin{array}{cc}
  1 & 0 \\
  0 & 1 \\
\end{array}%
\right), \quad B^{[k],1}_1 = \left(%
\begin{array}{cc}
  0 & \sqrt{\frac{2}{3}} \\
  0 & 0 \\
\end{array}%
\right), \quad B^{[k],2}_1 = \left(%
\begin{array}{cc}
  -\frac{1}{\sqrt{3}} & 0 \\
  0 & \frac{1}{\sqrt{3}} \\
\end{array}%
\right), \quad  B^{[k],3}_1 = \left(%
\begin{array}{cc}
  0 & 0 \\
  -\sqrt{\frac{2}{3}} & 0 \\
\end{array}%
\right).
\end{equation}

\noindent Note that $M^{ii'} = B^{[k],i}_1 \otimes
\overline{B^{[k],i'}_1}$. The transfer matrix reads
\begin{equation}
T = \sum_{i} M^{ii} = \frac{1}{3} \left(%
\begin{array}{cccc}
  1 & 0 & 0 & 2 \\
  0 & -1 & 0 & 0 \\
  0 & 0 & -1 & 0 \\
  2 & 0 & 0 & 1 \\
\end{array}%
\right)
\end{equation}

\noindent and has eigenvalues $1$ (of multiplicity $1$) and
$-\frac{1}{3}$ (of multiplicity $3$). The two-spin reduced density
matrix reads
\begin{equation} \label{rho-1-m}
\varrho_{1m} = \frac{1}{3} I \otimes \frac{1}{3} I + \left(
-\frac{1}{3} \right)^{m} \left( J_x \otimes J_x + J_y \otimes J_y
+ J_z \otimes J_z \right),
\end{equation}

\noindent where $J_{\alpha}$ is an operator for the spin
projection (in units of $\hbar$) on the $\alpha$ direction,
$\alpha = x,y,z$, see Eq.~\eqref{J-formula}.

Substituting Eq.~\eqref{rho-1-m} in Eq.~\eqref{K-m-through-H}, we
get
\begin{equation}
K_m[\varrho_S] = 4 \left(-\frac{1}{3}\right)^{m+1} g^2 \tau \left[
\frac{\sigma_x - \sigma_z}{\sqrt{2}}  \varrho_S \frac{\sigma_x -
\sigma_z}{\sqrt{2}} - \varrho_S \right].
\end{equation}

\noindent On the other hand, $K_m[\varrho_S] = \sum_{j}
(\lambda_j)^m L_{\rm nonlocal}^{(j)}[\varrho_S]$, i.e., in our
case we have a single contribution with $\lambda = -\frac{1}{3}$
and
\begin{equation}
L_{\rm nonlocal}[\varrho_S] = -\frac{4}{3} \left( \frac{\sigma_x -
\sigma_z}{\sqrt{2}}  \varrho_S \frac{\sigma_x -
\sigma_z}{\sqrt{2}} - \varrho_S \right).
\end{equation}

Finally, Eq.~\eqref{master-equation-stroboscopic} gives the
explicit master equation in the stroboscopic limit
\begin{equation} \label{equation-gksl}
\frac{d\varrho_S(t)}{dt} = \frac{g^2\tau}{3} \sum_{j=1,2,3} \left(
\sigma_j \varrho_S(t) \sigma_j - \varrho_S(t) \right) - \frac{g^2
\tau}{3}  \left( \frac{\sigma_x - \sigma_z}{\sqrt{2}} \varrho_S(t)
\frac{\sigma_x - \sigma_z}{\sqrt{2}} - \varrho_S(t) \right).
\end{equation}

\noindent The reader may notice the formally negative rate
$-\frac{g^2 \tau}{3}$ in front of the second dissipator term;
however, the total generator does have the
Gorini-Kossakowski-Sudarshan-Lindblad
form~\cite{gks-1976,lindblad-1976} because the Kossakowski matrix
is positive semidefinite. Therefore, the actual relaxation rates
are positive. The effect of the formally negative rate $-\frac{g^2
\tau}{3}$ in front of the second dissipator term in
Eq.~\eqref{equation-gksl} is that correlations among ancillas slow
down the relaxation as compared to the case of uncorrelated
ancillas (equation $\frac{d\varrho_S(t)}{dt} = L_{\rm
local}[\varrho_S(t)]$). Figure~\ref{figure10} illustrates this
phenomenon. Figure~\ref{figure10} also shows a good agreement
between the exact system dynamics and the system dynamics in the
stroboscopic limit and emphasizes the role of correlations.
Disregard of correlations among ancillas leads to a wrong result
(see the dashed line in Figure~\ref{figure10}).
\hfill$\triangle$\end{example}

\section{Effect of multipoint correlations in the higher-order stroboscopic limit}

We start with an example stimulating the discussion of the
higher-order stroboscopic limit.

\begin{example} \label{example-aklt-2}
Consider a qubit system interacting with an infinite chain of
spin-1 particles (ancillas) in the AKLT state as in
Example~\ref{example-aklt-1} with the only difference that the
system-ancilla interaction Hamiltonian now reads
\begin{equation} \label{H-aklt-2}
\hbar g H = \frac{\hbar g}{2} \sum_{j=x,y,z} \sigma_j \otimes J_j.
\end{equation}

\noindent For such an interaction $\braket{H}_{\rm anc} = 0$, so
the use of Proposition~\ref{proposition-stroboscopic} is
justified. Following the lines of Example~\ref{example-aklt-1} we
similarly calculate $L_{\rm local}$ and $L_{\rm nonlocal}$ in the
first-order stroboscopic limit; however, this contributions cancel
each other so that the right hand side of
Eq.~\eqref{master-equation-stroboscopic} vanishes in the
first-order stroboscopic limit and we get
$\frac{d\varrho_S(t)}{dt} = 0$.

The exact treatment of the problem via Eq.~\eqref{system-dynamics}
yields the depolarizing system dynamics
\begin{equation} \label{aklt-exact}
\varrho_S(t) = q(t) \varrho_S(0) + [1-q(t)] \frac{1}{2} I
\end{equation}

\noindent with the depolarization function
\begin{eqnarray}
&& q(t) = \left( \frac{1}{2} + \frac{x}{z} \right) \left( \frac{y
+ z}{27} \right)^{t/\tau} + \left( \frac{1}{2} - \frac{x}{z}
\right) \left( \frac{y - z}{27} \right)^{t/\tau}, \\
&& x = 2 + 7 \cos \frac{3 g \tau}{2}, \quad y = 7 + 2 \cos \frac{3
g \tau}{2}, \quad z = 2 \sqrt{y^2 + 27 \sin^2 \frac{3g\tau}{2}}.
\end{eqnarray}

\noindent A feature of the depolarizing dynamical
map~\eqref{aklt-exact} is that it is neither completely positive
divisible nor positive divisible for all $g \tau \leq \frac{2}{3}
\arccos (- \frac{11}{16})$ because
\begin{equation}
q (\tau) \geq 0 \quad \text{and} \quad q(2\tau) - q (\tau) =
\frac{2^5 y}{3^6} \sin^2 \frac{3 g \tau}{4}
> 0,
\end{equation}

\noindent i.e., the image of the system Bloch ball shrinks after
the first collision and then expands after the second collision.
If $g \tau = 4 \pi m /3$, $m \in \mathbb{N}$, then the system
experiences no evolution, i.e., $\varrho_S(t) = \varrho_S(0)$. If
$g\tau = 2 \pi /3$, then the Bloch ball experiences partial
inversion with the scaling parameter $-\frac{5}{27}$ after each
collision. The latter dynamics is completely positive divisible
though.

If $g \tau \ll 1$, then $q(t) \approx \exp(-\frac{1}{8} g^4 \tau^3
t)$, so the characteristic frequency of the system dynamics is
$g^4 \tau^3$. The first-order stroboscopic limit is unable to
reproduce such a behaviour because it is only sensitive to rates
$\sim g^2\tau$. This example stimulates us to develop (to some
extent) the theory of the higher-order stroboscopic limit.
\hfill$\triangle$\end{example}

We will refer to the limit $g\tau \rightarrow 0$, $g^{n+1}\tau^n =
{\rm const}$ as the $n$-th order stroboscopic limit. Surely, the
expressions $g$, $g^2\tau$, \ldots, $g^{n}\tau^{n-1}$ explode in
this limit; however, if their contribution to the system dynamics
vanishes, then the limit is well defined. The higher-order
contributions $g^{n+2}\tau^{n+1}$, $g^{n+3}\tau^{n+2}$, \ldots
vanish in the $n$-th order stroboscopic limit. In
Eq.~\eqref{L-local} for $L_{\rm local}$ we should throw away the
exploding terms (as they will cancel other exploding terms from
the memory kernel) and vanishing terms. In the memory kernel
series expansion~\eqref{kernel-series} we should keep the term
${\cal K}^{(n+1)}$, which describes $(n+1)$-point correlations
among ancillas. Recalling the notation in
Eq.~\eqref{Phi-expansion}, we get, for instance, the following
expression for the third-order memory-kernel:

\begin{eqnarray}
{\cal K}_{km}^{(3)} [\varrho_S] &=& \sum_{i_{k+1},i'_{k+1},
i_{k-m+1},i'_{k-m+1}} C^{(2)}_{ i_{k-m+1} i'_{k-m+1} i_{k+1}
i'_{k+1} }  \left( \Phi^{[k+1],{(1)}}_{i_{k+1} i'_{k+1}} \circ
\Phi^{[k-m+1],{(2)}}_{i_{k-m+1} i'_{k-m+1}}[\varrho_S]  +
\Phi^{[k+1],{(2)}}_{i_{k+1} i'_{k+1}} \circ
\Phi^{[k-m+1],{(1)}}_{i_{k-m+1} i'_{k-m+1}}[\varrho_S]
\right) \nonumber\\
&& + \sum_{l = k-m+2}^{k} \ \sum_{i_{k+1},i'_{k+1},i_l,i'_l,
i_{k-m+1},i'_{k-m+1}} C^{(3)}_{ i_{k-m+1} i'_{k-m+1} i_{l} i'_{l}
i_{k+1} i'_{k+1} } \Phi^{[k+1],{(1)}}_{i_{k+1} i'_{k+1}} \circ
\Phi^{[l],{(1)}}_{i_{l} i'_{l}} \circ
\Phi^{[k-m+1],{(1)}}_{i_{k-m+1} i'_{k-m+1}}[\varrho_S],
\end{eqnarray}

\noindent where $C^{(3)}_{ i_{k-m+1} i'_{k-m+1} i_{l} i'_{l}
i_{k+1} i'_{k+1} }$ is a three-point correlation function, which
is determined by tensor diagrams similar to those in
Fig.~\ref{figure9}c and reads

\begin{eqnarray}
&& C^{(3)}_{ i_{k-m+1} i'_{k-m+1} i_{l} i'_{l} i_{k+1} i'_{k+1} }
= {\cal C} (\ket{i'_{k-m+1}}\bra{i_{k-m+1}}, \,
\ket{i'_{l}}\bra{i_{l}}, \, \ket{i'_{k+1}}\bra{i_{k+1}}),
\\
&& {\cal C} (O,O',O'') \equiv \braket{O O' O''}_{\rm anc} -
\braket{O O'}_{\rm anc} \braket{ O''}_{\rm anc}- \braket{O }_{\rm
anc} \braket{O' O''}_{\rm anc} + \braket{O}_{\rm anc}
\braket{O'}_{\rm anc} \braket{O''}_{\rm anc},
\end{eqnarray}

\noindent with ${\cal C} (O,O',O'')$ being the third-order
Waldenfelds cumulant~\cite{waldenfels-1973,hegerfeldt-1988}. In
the second-order stroboscopic limit, the expression $g^3\tau^2
{\cal K}_{km}^{(3)}$ reduces to the time-local generator
$g^3\tau^2 L_{\rm nonlocal}$, which contributes to the final GKSL
master equation $\frac{d\varrho_S(t)}{dt} = L_{\rm
local}[\varrho_S(t)] + g^3\tau^2 L_{\rm nonlocal}[\varrho_S(t)]$.
The higher-order Waldenfelds cumulants are expressed through the
lower-order ones~\cite{hegerfeldt-1988}, thus enabling one to
achieve a desired stroboscopic order. To correctly describe
evolution in Example~\ref{example-aklt-2} one needs to consider
the third-order stroboscopic limit.

To give a broader view on the achieved result, the language of
tensor networks enabled us to relate the memory kernel components
with the multipoint correlation functions of the special form (the
Waldenfelds cumulants). Multipoint correlations of orders $n$,
$n-1$, \ldots, $2$ determine the system dynamics in the $(n-1)$-th
order stroboscopic limit. Although multitime correlation functions
have been used in the theory of open quantum systems (see, e.g.,
~\cite{pechen-2006,teretenkov-2020,gherardini-2022}), here we have
explicitly demonstrated their origin in the collision model. We
believe that the tensor network representation opens an avenue for
a further analysis of the effect of multipoint correlations on the
collisional dynamics, e.g., Wick's theorem for matrix product
states~\cite{hubener-2013} can be of great use.

\section{Conclusions}

We presented a tensor network approach to challenges in both the
standard collision model and the collision model with correlated
ancillas. We showed that the system-ancilla interactions in the
standard collision model induce a correlated state of the system
and ancillas that is naturally described by a right-canonical MPS
(if the system and ancillas are initially in pure states) and a
right-canonical MPDO (if the system and ancillas are initially in
mixed states). Since the description of MPS and MPDO requires much
less parameters as compared to a general multipartite state, we
believe that the revealed representation can find applications in
many practically relevant problems, e.g., this representation can
allow one to go well beyond $12$ collisions in the numerical study
of quantum thermometry~\cite{seah-2019}. As far as initially
correlated ancillas are concerned, we reviewed the recently
proposed approach to the tensor network description of the system
dynamics (with the emphasis on the two-point correlations) and
generalized it to the case of multipartite correlations among
ancillas. We showed conditions under which the higher-order
stroboscopic limit is to be considered and how the Waldenfelds
cumulants contribute to the memory-kernel master equation in this
case.

\begin{acknowledgments}
The author is greatly thankful to Alexander E. Teretenkov for
bringing Refs.~\cite{waldenfels-1973,hegerfeldt-1988} to his
attention. The author thanks Alexander N. Pechen, Valentin A.
Zagrebnov, Martin Plenio, and Francesco Ciccarello for useful
comments.
\end{acknowledgments}



\begin{thebibliography}{999}

\bibitem{rau-1963}
Rau J. Relaxation phenomena in spin and harmonic oscillator
systems. {\em Phys. Rev.} {\bf 1963}, {\em 129}, 1880--1888.

\bibitem{nachtergaele-2008}
Nachtergaele, B.; Vershynina, A.; Zagrebnov, V.A. Non-Equilibrium
states of a photon cavity pumped by an atomic beam. {\em Ann.
Henri Poincar\'{e}} {\bf 2008}, {\em 15}, 213--262.

\bibitem{scarani-2002}
Scarani, V.; Ziman, M.; \v{S}telmachovi\v{c}, P.; Gisin, N.;
Bu\v{z}ek, V. Thermalizing quantum machines: Dissipation and
entanglement. {\em Phys. Rev. Lett.} {\bf 2002}, {\em 88}, 097905.

\bibitem{ziman-osid-2002}
Ziman, M.; \v{S}telmachovi\v{c}, P.; Bu\v{z}ek, V. Description of
quantum dynamics of open systems based on collision-like models.
{\em Open Systems and Information Dynamics} {\bf 2005}, {\em 12},
81--91.

\bibitem{ziman-2005}
Ziman, M.; Bu\v{z}ek, V. All (qubit) decoherences: Complete
characterization and physical implementation. {\em Phys. Rev. A}
{\bf 2005}, {\em 72}, 022110.

\bibitem{grimmer-2016}
Grimmer, D.; Layden, D.; Mann, R.B.; Mart{\i}n-Mart{\i}nez, E.
Open dynamics under rapid repeated interaction. {\em Phys. Rev. A}
{\bf 2016}, {\em 94}, 032126.

\bibitem{ziman-2002}
Ziman, M.; \v{S}telmachovi\v{c}, P.; Bu\v{z}ek, V.; Hillery, M.;
Scarani, V.; Gisin, N. Diluting quantum information: An analysis
of information transfer in system-reservoir interactions. {\em
Phys. Rev. A} {\bf 2002}, {\em 65}, 042105.

\bibitem{ziman-2011}
Ziman, M.; Bu\v{z}ek, V. Open system dynamics of simple collision
models. In {\em Quantum Dynamics and Information}; Olkiewicz, R.,
Ceg{\l}a, W., Frydryszak, A., Garbaczewski, P., Jak\'{o}bczyk, L.,
Eds.; World Scientific: Singapore, 2011; pp. 199--227.

\bibitem{karevski-2009}
Karevski, D.; Platini, T. Quantum nonequilibrium steady states
induced by repeated interactions. {\em Phys. Rev. Lett.} {\bf
2009}, {\em 102}, 207207.

\bibitem{filip-2021}
Rom\'an-Ancheyta, R.; Kol\'a\ifmmode \check{r}\else \v{r}\fi{},
M.; Guarnieri, G.; Filip, R. Enhanced steady-state coherence via
repeated system-bath interactions. {\em Phys. Rev. A} {\bf 2021},
{\em 104}, 062209.

\bibitem{heineken-2021}
Heineken, D.; Beyer, K.; Luoma, K.; Strunz, W.T.
Quantum-memory-enhanced dissipative entanglement creation in
nonequilibrium steady states. {\em Phys. Rev. A} {\bf 2021}, {\em
104}, 052426.

\bibitem{daryanoosh-2018}
Daryanoosh, S.; Baragiola, B.Q.; Guff, T.; Gilchrist, A. Quantum
master equations for entangled qubit environments. {\em Phys. Rev.
A} {\bf 2018}, {\em 98}, 062104.

\bibitem{cakmak-2019}
\ifmmode \mbox{\c{C}}\else \c{C}\fi{}akmak, B.; Campbell, S.;
Vacchini, B.; M\"ustecapl\ifmmode \imath \else \i \fi{}o\ifmmode
\breve{g}\else \u{g}\fi{}lu, \"O.E.; Paternostro, M. Robust
multipartite entanglement generation via a collision model. Phys.
Rev. A {\bf 2019}, {\em 99}, 012319.

\bibitem{attal-2006}
Attal, S.; Pautrat, Y. From repeated to continuous quantum
interactions. {\em Ann. Henri Poincar\'{e}} {\bf 2006}, {\em 7},
59.

\bibitem{attal-2007}
Attal, S.; Joye, A. Weak coupling and continuous limits for
repeated quantum interactions. {\em J. Stat. Phys.} {\bf 2007},
{\em 126}, 1241--1283.

\bibitem{vargas-2008}
Vargas, R. Repeated interaction quantum systems: Van Hove limits
and asymptotic states. {\em J. Stat. Phys.} {\bf 2008}, {\em 133},
491--511.

\bibitem{li-2018}
Li, L.; Hall, M.J.W.; Wiseman, H.M. Concepts of quantum
non-Markovianity: A hierarchy. {\em Phys. Rep.} {\bf 2018}, {\em
759}, 1.

\bibitem{attal-2012}
Attal, S.; Petruccione, F.; Sinayskiy, I. Open quantum walks on
graphs. {\em Phys. Lett. A} {\bf 2012} {\em 376}, 1545.

\bibitem{attal-petruccione-2012}
Attal, S.; Petruccione, F.; Sabot, C.; Sinayskiy, I. Open quantum
random walks. {\em J. Stat. Phys.} {\bf 2012} {\em 147}, 832.

\bibitem{pellegrini-2014}
Pellegrini, C. Continuous time open quantum random walks and
non-Markovian Lindblad master equations. {\em J. Stat. Phys.} {\bf
2014}, {\em 154}, 838--865.

\bibitem{sinayskiy-2015}
Sinayskiy, I.; Petruccione, F. Microscopic derivation of open
quantum walks. {\em Phys. Rev. A} {\bf 2015}, 92, 032105.

\bibitem{liu-2017}
Liu, C.; Balu, R. Steady states of continuous-time open quantum
walks. {\em Quantum Inf. Process.} {\bf 2017}, 16, 173.

\bibitem{chia-2017}
Chia, A.; Paterek, T.; Kwek, L.C. Hitting statistics from quantum
jumps. {\em Quantum} {\bf 2017}, {\em 1}, 19.

\bibitem{bruneau-2014}
Bruneau, L.; Joye, A.; Merkli, M. Repeated interactions in open
quantum systems. {\em J. Math. Phys.} {\bf 2014}, {\em 55},
075204.

\bibitem{bruneau-2006}
Bruneau, L.; Joye, A.; Merkli, M. Asymptotics of repeated
interaction quantum systems. {\em Journal of Functional Analysis}
{\bf 2006}, {\em 239}, 310--344.

\bibitem{tamura-2016}
Tamura, H.; Zagrebnov, V.A. Dynamics of an open system for
repeated harmonic perturbation. {\em J. Stat. Phys.} {\bf 2016},
{\em 163}, 844--867.

\bibitem{bruneau-2008}
Bruneau, L.; Joye, A.; Merkli, M. Random repeated interaction
quantum systems. {\em Commun. Math. Phys.} {\bf 2008}, {\em 284},
553--581.

\bibitem{nechita-2012}
Nechita, I; Pellegrini, C. Random repeated quantum interactions
and random invariant states. {\em Probab. Theory Relat. Fields}
{\bf 2012}, {\em 152}, 299--320.


\bibitem{purkayastha-2021}
Purkayastha, A.; Guarnieri, G.; Campbell, S.; Prior, J.; Goold, J.
Periodically refreshed baths to simulate open quantum many-body
dynamics. {\em Phys. Rev. B} {\bf 2021}, {\bf 104}, 045417.

\bibitem{cattaneo-2021}
Cattaneo, M.; De Chiara, G.; Maniscalco, S.; Zambrini, R.; Giorgi,
G.L. Collision models can efficiently simulate any multipartite
Markovian quantum dynamics. {\em Phys. Rev. Lett.} {\bf 2021},
{\em 126}, 130403.

\bibitem{garcia-perez-2020}
Garc\'{\i}a-P\'{e}rez, G.; Rossi, M.A.C.; Maniscalco, S. IBM Q
Experience as a versatile experimental testbed for simulating open
quantum systems. {\em npj Quantum Inf.} {\bf 2020} {\em 6}, 1.

\bibitem{filippov-2020}
Filippov, S.N.; Semin, G.N.; Pechen, A.N. Quantum master equations
for a system interacting with a quantum gas in the low-density
limit and for the semiclassical collision model. {\em Phys. Rev.
A} {\bf 2020}, {\em 101}, 012114.

\bibitem{kosloff-2019}
Kosloff, R. Quantum thermodynamics and open-systems modeling. {\em
J. Chem. Phys.} {\bf 2019}, {\em 150}, 204105.

\bibitem{seah-2019}
Seah, S.; Nimmrichter, S.; Grimmer, D.; Santos, J.P.; Scarani, V.;
Landi, G.T. Collisional quantum thermometry. {\em Phys. Rev.
Lett.} {\bf 2019}, {\em 123}, 180602.

\bibitem{strasberg-2019}
Strasberg, P. Repeated interactions and quantum stochastic
thermodynamics at strong coupling. {\em Phys. Rev. Lett.} {\bf
2019}, {\em 123}, 180604.

\bibitem{pichler-2016}
Pichler, H.; Zoller, P. Photonic circuits with time delays and
quantum feedback. {\em Phys. Rev. Lett.} {\bf 2016}, {\em 116},
093601.

\bibitem{guimond-2017}
Guimond, P.-O.; Pletyukhov, M.; Pichler, H.; Zoller, P. Delayed
coherent quantum feedback from a scattering theory and a matrix
product state perspective. Quantum Sci. Technol. {\bf 2017}, {\em
2}, 044012.

\bibitem{ciccarello-2017}
Ciccarello, F. Collision models in quantum optics. {\em Quantum
Measurements and Quantum Metrology} {\bf 2017}, {\em 4}, 53--63.

\bibitem{gross-2018}
Gross, J.A.; Caves, C.M.; Milburn, G.J.; Combes, J. Qubit models
of weak continuous measurements: Markovian conditional and
open-system dynamics. {\em Quantum Sci. Technol.} {\bf 2018}, {\em
3}, 024005.

\bibitem{fisher-2018}
Fischer, K.A.; Trivedi, R.; Ramasesh, V.; Siddiqi, I.;
Vu{\v{c}}kovi{\'{c}}, J. Scattering into one-dimensional
waveguides from a coherently-driven quantum-optical system. {\em
Quantum} {\bf 2018}, {\em 2}, 69.

\bibitem{cilluffo-2020}
Cilluffo, D.; Carollo, A.; Lorenzo, S.; Gross, J.A.; Palma, G.M.;
Ciccarello, F. Collisional picture of quantum optics with giant
emitters. {\em Phys. Rev. Research} {\bf 2020} {\em 2}, 043070.

\bibitem{carmele-2020}
Carmele, A.; Nemet, N.; Canela, V.; Parkins, S. Pronounced
non-Markovian features in multiply excited, multiple emitter
waveguide QED: Retardation induced anomalous population trapping.
{\em Phys. Rev. Research} {\bf 2020}, {\em 2}, 013238.

\bibitem{ferreira-2021}
Ferreira, V.S.; Banker, J.; Sipahigil, A.; Matheny, M.H.; Keller,
A.J.; Kim, E.; Mirhosseini, M.; Painter, O. Collapse and revival
of an artificial atom coupled to a structured photonic reservoir.
{\em Phys. Rev. X} {\bf 2021}, {\em 11}, 041043.

\bibitem{wein-2021}
Wein, S.C.; Loredo, J.C.; Maffei, M.; Hilaire, P.; Harouri, A.;
Somaschi, N.; Lema\^itre, A.; Sagnes, I.; Lanco, L.; Krebs, O.;
Auff\`eves, A.; Simon, C.; Senellart, P.; Ant\'on-Solanas, C.
Photon-number entanglement generated by sequential excitation of a
two-level atom. Available online: https://arxiv.org/abs/2106.02049
(accessed on 21 February 2022).

\bibitem{maffei-2022}
Maffei, M.; Camati, P.A.; Auff\`eves, A. Closed-system solution of
the 1D atom from collision model. {\em Entropy} {\bf 2022}, {\em
24}, 151.

\bibitem{gheri-1998}
Gheri, K.M.; Ellinger, K.; Pellizzari, T.; Zoller, P.
Photon-wavepackets as flying quantum bits. {\em Fortschr. Phys.}
{\bf 1998}, {\em 46},401--415.

\bibitem{baragiola-2012}
Baragiola, B.Q.; Cook, R.L.; Bra\'{n}czyk, A.M.; Combes, J.
N-photon wave packets interacting with an arbitrary quantum
system. {\em Phys. Rev. A} {\bf 2012}, {\em 86}, 013811.

\bibitem{dabrowska-2020}
D\k{a}browska, A.M. From a posteriori to a priori solutions for a
two-level system interacting with a single-photon wavepacket. {\em
J. Opt. Soc. Am. B} {\bf 2020}, {\em 37}, 1240--1248.

\bibitem{dabrowska-2021}
D\k{a}browska, A.; Chru\'{s}ci\'{n}ski, D.; Chakraborty, S.;
Sarbicki, G. Eternally non-Markovian dynamics of a qubit
interacting with a single-photon wavepacket. {\em New J. Phys.}
{\bf 2021}, {\em 23}, 123019.

\bibitem{rybar-2012}
Ryb\'{a}r, T.; Filippov, S.N.; Ziman, M.; Bu\v{z}ek, V. Simulation
of indivisible qubit channels in collision models. {\em J. Phys.
B: At. Mol. Opt. Phys.} {\bf 2012}, {\em 45}, 154006.

\bibitem{filippov-2017}
Filippov, S.N.; Piilo, J.; Maniscalco, S.; Ziman, M. Divisibility
of quantum dynamical maps and collision models. {\em Phys. Rev. A}
{\bf 2017}, {\em 96}, 032111.

\bibitem{ciccarello-pra-2013}
Ciccarello, F.; Palma, G.M.; Giovannetti, V. Collision-model-based
approach to non-Markovian quantum dynamics. {\em  Phys. Rev. A}
{\bf 2013}, {\em 87}, 040103(R).

\bibitem{ciccarello-ps-2013}
Ciccarello, F.; Giovannetti, V. A quantum non-Markovian collision
model: incoherent swap case. {\em Phys. Scr.} {\bf 2013}, {\em
T153}, 014010.

\bibitem{kretschmer-2016}
Kretschmer, S.; Luoma, K.; Strunz, W. T. Collision model for
non-Markovian quantum dynamics. {\em Phys. Rev. A} {\bf 2016},
{\em 94}, 012106.

\bibitem{campbell-2018}
Campbell, S.; Ciccarello, F.; Palma, G.M.; Vacchini, B.
System-environment correlations and Markovian embedding of quantum
non-Markovian dynamics. {\em Phys. Rev. A} {\bf 2018}, {\em 98},
012142.

\bibitem{lorenzo-2017}
Lorenzo, S.; Ciccarello, F.; Palma, G.M. Composite quantum
collision models. {\em Phys. Rev. A} {\bf 2017}, {\em 96}, 032107.

\bibitem{pellegrini-2009}
Pellegrini, C.; Petruccione, F. Non-Markovian quantum repeated
interactions and measurements. {\em J. Phys. A: Math. Theor.} {\bf
2009}, {\em 42}, 425304.

\bibitem{cilluffo-2019}
Cilluffo, D.; Ciccarello, F. Quantum non-Markovian collision
models from colored-noise baths. In {\em Advances in Open Systems
and Fundamental Tests of Quantum Mechanics, Springer Proceedings
in Physics, vol. 237}; Vacchini, B., Breuer, H.-P., Bassi, A.,
Eds.; Springer, Cham, 2019; pp. 29--40.

\bibitem{taranto-2019}
Taranto, P.; Milz, S.; Pollock, F.A.; Modi, K. Structure of
quantum stochastic processes with finite Markov order. {\em Phys.
Rev. A} {\bf 2019}, {\em 99}, 042108.

\bibitem{kretschmann-2005}
Kretschmann, D.; Werner, R.F. Quantum channels with memory. {\em
Phys. Rev. A} {\bf 2005}, {\em 72}, 062323.

\bibitem{plenio-2007}
Plenio, M.B.; Virmani, S. Spin chains and channels with memory.
{\em Phys. Rev. Lett.} {\bf 2007}, {\em 99}, 120504.

\bibitem{plenio-2008}
Plenio, M.B.; Virmani, S. Many-body physics and the capacity of
quantum channels with memory. {\em New J. Phys.} {\bf 2008}, {\em
10}, 043032.

\bibitem{rybar-2008}
Ryb\'{a}r, T.; Ziman, M. Repeatable quantum memory channels. {\em
Phys. Rev. A} {\bf 2008}, {\em 78}, 052114.

\bibitem{rybar-2009}
Ryb\'{a}r, T.; Ziman, M. Quantum finite-depth memory channels:
Case study. {\em Phys. Rev. A} {\bf 2009}, {\em 80}, 042306.

\bibitem{giovannetti-2012}
Giovannetti, V.; Palma, G.M. Master equations for correlated
quantum channels. {\em Phys. Rev. Lett.} {\bf 2012}, {\em 108},
040401.

\bibitem{palma-2012}
Giovannetti, V.; Palma, G.M. Master equation for cascade quantum
channels: a collisional approach. {\em J. Phys. B: At. Mol. Opt.
Phys.} {\bf 2012}, {\em 45}, 154003.

\bibitem{rybar-2015}
Ryb\'{a}r, T.; Ziman, M. Process estimation in the presence of
time-invariant memory effects. {\em Phys. Rev. A} {\bf 2015}, {\em
92}, 042315.

\bibitem{ciccarello-2022}
Ciccarello, F.; Lorenzo, S.; Giovannetti, V.; Palma, G.M. Quantum
collision models: Open system dynamics from repeated interactions.
{\em Physics Reports} {\bf 2022}, {\em 954}, 1--70.

\bibitem{campbell-2021}
Campbell, S.; Vacchini, B. Collision models in open system
dynamics: A versatile tool for deeper insights? {\em EPL} {\bf
2021}, {\em 133}, 60001.

\bibitem{aklt-1987}
Affleck, I.; Kennedy, T.; Lieb, E.H.; Tasaki, H. Rigorous results
on valence-bond ground states in antiferromagnets. {\em Phys. Rev.
Lett.} {\bf 1987}, {\em 59}, 799.

\bibitem{comar-2021}
Comar, N.E.; Landi, G.T. Correlations breaking homogenization.
{\em Phys. Rev. A} {\bf 2021}, {\em 104}, 032217.

\bibitem{filippov-2022}
Filippov, S.N.; Luchnikov, I.A. Collisional open quantum dynamics
with a generally correlated environment: Exact solvability in
tensor networks. Available online:
https://arxiv.org/abs/2202.04697 (accessed on 21 February 2022).

\bibitem{perez-garcia-2007}
P\'{e}rez-Garc\'{\i}a, D.; Verstraete, F.; Wolf, M.M.; Cirac, J.I.
Matrix product state representations. {\em Quantum Information and
Computation} {\bf 2007}, {\em 7}, 401.

\bibitem{verstraete-2008}
Verstraete, F.; Murg, V.; Cirac, J.I. Matrix product states,
projected entangled pair states, and variational renormalization
group methods for quantum spin systems. {\em Advances in Physics}
{\bf 2008}, {\em 57}, 143.

\bibitem{schollwock-2011}
Schollw\"{o}ck, U. The density-matrix renormalization group in the
age of matrix product states. {\em Annals of Physics} {\bf 2011},
{\em 326}, 96.

\bibitem{cirac-2021}
Cirac, J.I.; P\'{e}rez-Garc\'{\i}a, D.; Schuch, N.; Verstraete, F.
Matrix product states and projected entangled pair states:
Concepts, symmetries, theorems. {\em Rev. Mod. Phys.} {\bf 2021},
{\em 93}, 045003.

\bibitem{orus-2014}
Or\'{u}s, R. A practical introduction to tensor networks: Matrix
product states and projected entangled pair states. {\em Annals of
Physics} {\bf 2014}, {\em 349}, 117-158.

\bibitem{orus-2019}
Or\'{u}s, R. Tensor networks for complex quantum systems. {\em
Nature Reviews Physics} {\bf 2019}, {\em 1}, 538-550.

\bibitem{montangero-2018}
Montangero, S. \textit{Introduction to Tensor Network Methods};
Springer: New York, 2018.

\bibitem{vidal-2003}
Vidal, G. Efficient classical simulation of slightly entangled
quantum computations. {\em Phys. Rev. Lett.} {\bf 2003}, {\em 91},
147902.

\bibitem{luchnikov-qgopt-2021}
Luchnikov, I.A.; Ryzhov, A.; Filippov, S.N.; Ouerdane, H. QGOpt:
Riemannian optimization for quantum technologies. {\em SciPost
Phys.} {\bf 2021}, {\em 10}, 079.

\bibitem{luchnikov-2021}
Luchnikov, I.A.; Krechetov, M.E.; Filippov, S.N. Riemannian
geometry and automatic differentiation for optimization problems
of quantum physics and quantum technologies. {\em New J. Phys.}
{\bf 2021}, {\em 23}, 073006.

\bibitem{verstraete-2004}
Verstraete, F.; Garc\'{\i}a-Ripoll, J.J.; Cirac, J.I. Matrix
product density operators: Simulation of finite-temperature and
dissipative systems. {\em Phys. Rev. Lett.} {\bf 2004}, {\em 93},
207204.

\bibitem{zwolak-2004}
Zwolak, M.; Vidal, G. Mixed-state dynamics in one-dimensional
quantum lattice systems: A time-dependent superoperator
renormalization algorithm. {\em Phys. Rev. Lett.} {\bf 2004}, {\em
93}, 207205.

\bibitem{chen-2020}
Chen, C.-F.; Kato, K.; Brand\~{a}o, F.G.S.L. Matrix Product
Density Operators: when do they have a local parent Hamiltonian?
Available online: https://arxiv.org/abs/2010.14682 (accessed on 21
February 2022).

\bibitem{bondarenko-2021}
Bondarenko, D. Constructing k-local parent Lindbladians for matrix
product density operators. Available online:
https://arxiv.org/abs/2110.13134 (accessed on 21 February 2022).

\bibitem{wood-2015}
Wood, C.J.; Biamonte, J.D.; Cory, D.G. Tensor networks and
graphical calculus for open quantum systems. {\em Quantum
Information and Computation} {\bf 2015}, {\em 15}, 759-811.

\bibitem{dhand-2018}
Dhand, I.; Engelkemeier, M.; Sansoni, L.; Barkhofen, S.;
Silberhorn, C.; Plenio, M.B. Proposal for quantum simulation via
all-optically-generated tensor network states. {\em Phys. Rev.
Lett.} {\bf 2018}, {\em 120}, 130501.

\bibitem{lubash-2018}
Lubasch, M.; Valido, A.A.; Renema, J.J.; Kolthammer, W.S.; Jaksch,
D.; Kim, M.S.; Walmsley, I.; Garc{\i}a-Patr\'{o}n, R. Tensor
network states in time-bin quantum optics. {\em Phys. Rev. A} {\bf
2018}, {\bf 97}, 062304.

\bibitem{istrati-2020}
Istrati, D.; Pilnyak, Y.; Loredo, J.C.; Ant\'{o}n, C.; Somaschi,
N.; Hilaire, P.; Ollivier, H.; Esmann, M.; Cohen, L.; Vidro, L.;
Millet, C.; Lema\^{\i}tre, A.; Sagnes, I.; Harouri, A.; Lanco, L.;
Senellart, P.; Eisenberg, H.S. Sequential generation of linear
cluster states from a single photon emitter. {\em Nat. Commun.}
{\bf 2020}, {\em 11}, 5501.

\bibitem{besse-2020}
Besse, J.-C.; Reuer, K.; Collodo, M.C.; Wulff, A.; Wernli, L.;
Copetudo, A.; Malz, D.; Magnard, P.; Akin, A.; Gabureac, M.;
Norris, G.J.; Cirac, J.I.; Wallraff, A.; Eichler, C. Realizing a
deterministic source of multipartite-entangled photonic qubits.
{\em Nat. Commun.} {\bf 2020}, {\em 11}, 4877.

\bibitem{tiurev-2020}
Tiurev, K.; Appel, M.H.; Mirambell, P.L.; Lauritzen, M.B.;
Tiranov, A.; Lodahl, P.; S{\o}rensen, A.S. High-fidelity
multi-photon-entangled cluster state with solid-state quantum
emitters in photonic nanostructures. Available online:
https://arxiv.org/abs/2007.09295 (accessed on 21 February 2022).

\bibitem{wei-2021}
Wei, Z.-Y.; Malz, D.; Gonz\'alez-Tudela, A.; Cirac, J.I.
Generation of photonic matrix product states with Rydberg atomic
arrays. {\em Phys. Rev. Research} {\bf 2021}, {\em 3}, 023021.

\bibitem{dalzell-2019}
Dalzell, A.M.; Brand\~{a}o, F.G.S.L. Locally accurate MPS
approximations for ground states of one-dimensional gapped local
Hamiltonians. {\em Quantum} {\bf 2019}, {\em 3}, 187.

\bibitem{pollock-pra-2018}
Pollock, F.A.; Rodr\'{\i}guez-Rosario, C.; Frauenheim, T.;
Paternostro, M.; Modi, K. Non-Markovian quantum processes:
Complete framework and efficient characterization. {\em Phys. Rev.
A} {\bf 2018}, {\bf 97}, 012127.

\bibitem{pollock-prl-2018}
Pollock, F.A.; Rodr\'{\i}guez-Rosario, C.; Frauenheim, T.;
Paternostro, M.; Modi, K. Operational Markov condition for quantum
processes. {\em Phys. Rev. Lett.} {\bf 2018}, {\em 120}, 040405.

\bibitem{white-2020}
White, G.A.L.; Hill, C.D.; Pollock, F.A.; Hollenberg, L.C.L.;
Modi, K. Demonstration of non-Markovian process characterisation
and control on a quantum processor. {\em Nat. Commun.} {\bf 2020},
{\em 11}, 6301.

\bibitem{taranto-2020}
Taranto, P. Memory effects in quantum processes. {\em Int. J.
Quantum Inf.} {\bf 2020}, {\em 18}, 1941002.

\bibitem{luchnikov-2019}
Luchnikov, I.A.; Vintskevich, S.V.; Ouerdane, H.; Filippov, S.N.
Simulation complexity of open quantum dynamics: Connection with
tensor networks. {\em Phys. Rev. Lett.} {\bf 2019}, {\em 122},
160401.

\bibitem{luchnikov-2020}
Luchnikov, I.A.; Vintskevich, S.V.; Grigoriev, D.A.; Filippov,
S.N. Machine learning non-Markovian quantum dynamics. {\em Phys.
Rev. Lett.} {\bf 2020}, {\em 124}, 140502.

\bibitem{chruscinski-maniscalco-2014}
Chru\'{s}ci\'{n}ski, D.; Maniscalco, S. Degree of non-Markovianity
of quantum evolution. {\em Phys. Rev. Lett.} {\bf 2014}, {\em
112}, 120404.

\bibitem{fgl-2020}
Filippov, S.N.; Glinov, A.N.; Lepp\"{a}j\"{a}rvi, L. Phase
covariant qubit dynamics and divisibility. {\em Lobachevskii
Journal of Mathematics} {\bf 2020}, {\em 41}, 617--630.

\bibitem{breuer-2002}
Breuer, H.-P.; Petruccione, F. \emph{The Theory of Open Quantum
Systems}, chapter 9; Oxford University Press: Oxford, 2002.

\bibitem{nakajima-1958}
Nakajima, S. On quantum theory of transport phenomena: Steady
diffusion. {\em Prog. Theor. Phys.} {\bf 1958}, {\em 20},
948--959.

\bibitem{zwanzig-1960}
Zwanzig, R. Ensemble method in the theory of irreversibility. {\em
J. Chem. Phys.} {\bf 1960}, {\em 33}, 1338--1341.

\bibitem{luchnikov-2017}
Luchnikov, I.A.; Filippov, S.N. Quantum evolution in the
stroboscopic limit of repeated measurements. {\em Phys. Rev. A}
{\bf 2017}, {\em 95}, 022113.

\bibitem{grimaudo-2020}
Grimaudo, R.; Messina, A.; Sergi, A.; Vitanov, N.V.; Filippov,
S.N. Two-qubit entanglement generation through non-Hermitian
Hamiltonians induced by repeated measurements on an ancilla. {\em
Entropy} {\bf 2020}, {\em 22}, 1184.

\bibitem{chruscinski-2010}
Chru\'{s}ci\'{n}ski, D.; Kossakowski, A. Non-Markovian quantum
dynamics: Local versus nonlocal. {\em Phys. Rev. Lett.} {\bf
2010}, {\em 104}, 070406.

\bibitem{smirne-2010}
Smirne, A.; Vacchini, B. Nakajima-Zwanzig versus
time-convolutionless master equation for the non-Markovian
dynamics of a two-level system. {\em Phys. Rev. A} {\bf 2010},
{\em 82}, 022110.

\bibitem{filippov-2018}
Filippov, S.N.; Chru\'{s}ci\'{n}ski, D. Time deformations of
master equations. {\em Phys. Rev. A} {\bf 2018}, {\em 98}, 022123.

\bibitem{gks-1976}
Gorini, V.; Kossakowski, A.; Sudarshan, E.C.G. Completely positive
dynamical semigroups of n-level systems. {\em J. Math. Phys.} {\bf
1976}, {\bf 17}, 821.

\bibitem{lindblad-1976}
Lindblad, G. On the generators of quantum dynamical semigroups.
Comm. Math. Phys. {\bf 1976}, {\em 48}, 119.

\bibitem{waldenfels-1973}
von Waldenfels, W. An approach to the theory of pressure
broadening of spectral lines. In {\em Probability and information
theory II}; Behara, M., Krickeberg, K., Wolfowitz, J., Eds.;
Springer: Berlin, Heidelberg, 1973; pp. 19--69.

\bibitem{hegerfeldt-1988}
Hegerfeldt, G.C.; Schulze, H. Noncommutative cumulants for
stochastic differential equations and for generalized Dyson
series. {\em Journal of Statistical Physics} {\bf 1988}, {\em 51},
691--710.

\bibitem{pechen-2006}
Pechen, A.N. The multitime correlation functions, free white
noise, and the generalized Poisson statistics in the low density
limit. {\em J. Math. Phys.} {\bf 2006}, {\em 47}, 033507.

\bibitem{teretenkov-2020}
Nosal', I.A.; Teretenkov, A.E. Exact dynamics of moments and
correlation functions for GKSL fermionic equations of Poisson
type. {\em Math Notes} {\bf 2020}, {\em 108}, 911--915.

\bibitem{gherardini-2022}
Gherardini, S.; Smirne, A.; Huelga, S.F.; Caruso, F.
Transfer-tensor description of memory effects in open-system
dynamics and multi-time statistics. {\em Quantum Sci. Technol.}
{\bf 2022}, {\em 7} 025005.

\bibitem{hubener-2013}
H\"{u}bener, R.; Mari, A.; Eisert, J. Wick's theorem for matrix
product states. {\em Phys. Rev. Lett.} {\bf 2013}, {\em 110},
040401.

\end{thebibliography}
\end{document}